\documentclass[pra,showpacs,twocolumn,superscriptaddress,floatfix,aps]{revtex4-2}
\usepackage{amsmath,amssymb,amsfonts,bm}
\usepackage{braket}
\usepackage{float}
\usepackage{graphicx}
\usepackage{physics}
\usepackage{mathtools}
\usepackage{hyperref}
\usepackage{comment}
\usepackage[normalem]{ulem}
\usepackage[T1]{fontenc}
\hypersetup{colorlinks,%
    linkcolor=blue,%
    citecolor=blue,%
    urlcolor=blue}
\usepackage{color}
\usepackage{tikz}
\usetikzlibrary{shapes}

\begin{document}

\title{Vortex configuration dependent equilibrium and non-equilibrium states in two-dimensional quantum turbulence}

\author{Shawan K. Jha}
\affiliation{Department of Physics, Indian Institute of Technology Guwahati, Guwahati 781039, Assam, India}
\author{Makoto Tsubota}
\affiliation{Department of Physics, Nambu Yoichiro Institute of Theoretical and Experimental Physics (NITEP), Osaka Metropolitan University, Sumiyoshi-Ku, Osaka 558-8585, Japan}
\author{Pankaj K. Mishra}
\affiliation{Department of Physics, Indian Institute of Technology Guwahati, Guwahati 781039, Assam, India}
\date{\today}
 \begin{abstract}
In this work, we analyze the evolution of four vortex configurations, namely, dipole, plasma, cluster, and lattice, using the two-dimensional mean-field Gross-Pitaevskii equation, focusing on their dynamical decay and approach to the equilibrium. Our analysis reveals that the cluster vortex configuration reaches equilibrium more rapidly than the others, while the dipole, plasma, and lattice configurations exhibit persistent non-equilibrium behavior, tending toward non-thermal fixed points. Specifically, the cluster configuration follows Kolmogorov-like scaling ($\varepsilon^{i}(k)\sim k^{-5/3}$) in the incompressible spectrum, while the other configurations follow Vinen-like scaling ($\varepsilon^{i}(k)\sim k^{-1}$). In the compressible spectrum, the cluster case exhibits a $k$ scaling, indicating full mode equilibration, while for the other configurations, the modes thermalize only above a critical wave number. Additionally, the transfer function for the cluster configuration displays a Gaussian distribution, typical of equilibrium states, while the other configurations exhibit skewed Gaussian or exponential distributions, indicative of their non-equilibrium nature. Finally, the particle number spectra show that the cluster case follows dynamical scaling closer to equilibrium, while the dipole, plasma, and lattice configurations evolve towards non-thermal fixed points. Our findings provide new insights into the dynamics of vortex configurations and their approach to equilibrium or non-equilibrium states, offering guidance for future studies on quantum turbulence and its control.
 \end{abstract}
\maketitle
\section{Introduction}
Turbulence remains one of the most challenging and unresolved problems in classical physics, characterized by chaotic, multiscale flow dynamics and nonlinear interactions~\cite{Frisch1995,Barenghi:CUP2023}. The complexity lies in the vast number of degrees of freedom involved, with turbulent flow consisting of a range of interacting eddies and vortices that transfer energy nonlinearly across scales. This energy cascade, particularly in high Reynolds number flows, complicates predictive modeling. Despite advances in statistical models and computational approaches, a universally accepted theory that captures turbulence in all its forms remains elusive~\cite{Sreenivasan:ARCMP2025}.

In recent years, quantum fluids have emerged as a compelling platform to investigate turbulent phenomena under more controlled and quantifiable conditions~\cite{Nazarenko:JLTP2006, Barenghi:CUP2023, Tsubota:OUP2025, Madeira:AVSQS2020}. Systems such as superfluid helium and ultracold Bose–Einstein condensates (BECs) offer unique advantages for studying turbulence. In contrast to classical fluids, quantum fluids exhibit quantized circulation, restricted to discrete values of $h/m$ where $h$ is Planck's constant and $m$ is the mass of the Boson. This leads to the formation of discrete quantum vortices, each carrying a single quantum of circulation~\cite{Donnelly1991}. Furthermore, quantum fluids are inviscid, eliminating classical dissipative mechanisms and enabling a more direct investigation of the conservative energy transfer dynamics inherent in turbulence.

The study of quantum turbulence in BECs has advanced rapidly in both theoretical~\cite{Nazarenko:PDNP2006, Proment:PDNP2012, Proment:PRA2009, Bradley:PRX2012, Reeves:PRL2013, Billam:PRA2015, Groszek:PRA2018, Barenghi:PNAS2014, Jha:PRF2025} and experimental~\cite{  Henn:PRL2009, Kwon:PRA2014, Navon:Sci2019, Gauthier:Sci2019, Johnstone:Sci2019} contexts. These investigations have revealed a range of phenomena including Kolmogorov-like scaling in kinetic energy spectra~\cite{Nore:PRL1997, Kobayashi:PRL2005, Bradley:PRX2012}, vortex clustering~\cite{White:PRL2010, White:PRA2012, Neely:PRL2013, Stagg:PRA2015}, and non-thermal fixed points (NTFPs)~\cite{Berges:NPB2009, Nowak:PRB2011, Orioli:PRD2015} which represent states where the system exhibits universal scaling behavior far from thermal equilibrium.

In two-dimensional quantum turbulence, a system can evolve towards an equilibrium or non-equilibrium state, depending on its initial conditions, such as a rapid cooling quench from a highly out-of-equilibrium configuration~\cite{Berges:PRL2012, Nowak:NJP2014}. Quenching triggers complex bidirectional transport processes: an inverse cascade of momenta from intermediate to lower momenta, alongside a direct cascade of energy from intermediate to higher wave numbers. These processes are key to understanding the evolution of vortex and antivortex configurations in the turbulent flow. As the system approaches a NTFP, the momentum distribution of the system, especially in a BEC, exhibits self-similarity and power-law scaling over a specific momentum range. The scaling exponent in this regime provides valuable insight into the nature of the underlying transport mechanisms and helps classify the system's evolution within a particular universality class. Despite this, a comprehensive classification of all possible universality classes in such non-thermal dynamics remains a major open problem~\cite{Orioli:PRD2015, Chantesana:PRA2019, Mikheev:EPJ2023, Madeira:PNAS2024}. Understanding these classes is critical for developing predictive models of quantum turbulence and extending the concept of NTFPs to systems beyond BECs, such as other quantum fluids with complex vortex-antivortex dynamics.

A central question in quantum turbulence is how systems relax toward equilibrium starting from highly non-equilibrium initial conditions. Different theoretical perspectives have addressed this. Some theories define equilibrium as the decay of all vortices, leaving only compressible excitations (phonons) exhibiting thermal scaling~\cite{Numasato:PRA2010}, while others focus on the statistical evolution of vortex positions and charges~\cite{Reeves:PRX2022, Valani:NJOP2018, Johnstone:Sci2019}. However, the full understanding of the coupled dynamics between the condensate and vortex configurations is still incomplete~\cite{Tsubota:OUP2025}. In particular, the exchange of energy between compressible and incompressible components adds richness to the system's evolution and complicates the identification of true thermodynamic equilibrium.

In this work, we aim to investigate how different initial configurations of vortices and antivortices evolve over time in a two-dimensional turbulent BEC. Specifically, we focus on whether such systems evolve toward well-defined quasi-equilibrium states, and how energy and particle number redistribute across spatial scales during this evolution. We consider four prototypical non-equilibrium vortex configurations: A vortex dipole gas, consisting of randomly oriented and positioned vortex dipoles; A random vortex gas, with equal numbers of vortices and antivortices placed randomly throughout the condensate; A clustered configuration, with spatially separated vortex and antivortex clusters; A square vortex lattice, where vortices are arranged in a regular pattern with alternating charges.

Each vortex configuration is evolved using the Gross–Pitaevskii Equation (GPE), which captures the mean-field dynamics of dilute, weakly interacting Bose gases at zero temperature. To characterize the evolution, we analyze the incompressible kinetic energy spectra, particle number spectra, and define a particle number transfer function. These tools allow us to quantify how particles and energy are redistributed across scales as the system evolves.

Our work unfolds as follows. In Sec.~\ref{sec:model} we present the details of mean-field Gross-Pitaevskii model and the diagnostic measures used to characterize vortex dynamics. Sec.~\ref{sec:numresults} presents the numerical results for the four initial vortex configurations, including vortex-number decay, vortex sign correlations, energy evolution, and spectral analyses of kinetic energy and particle-number transport. Finally in Sec.~\ref{sec:con} we summarize our main findings and discuss several directions for future research.

\section{Governing equations and details of global quantities used to characterize the dynamical evolution of the vortex configuration}
\label{sec:model}
The behavior of two dimensional BECs, with tight confinement along one axis, can be described at absolute temperature using the Gross-Pitaevskii equation (GPE) which is given by~\cite{Pethick:CUP2008, Pitaevskii:OUP2016}
\begin{equation}\label{eq:dimgpe}
    i\hbar\partial_t\Psi(\vb{r}) = -\frac{\hbar^2}{2m}\vb{\nabla}^2\Psi(\vb{r}) + g|\Psi(\vb{r})|^2\psi(\vb{r}),
\end{equation}
where $\Psi(\vb{r})$ denotes the macroscopic wavefunction and $g$ represents the effective interaction strength in two dimensions. In order to simplify the analysis, we rescale the equation by introducing dimensionless variables. Specifically, we define length in units of the healing length $\xi = \hbar/\sqrt{m\mu}$, time in terms of $\tau = \hbar/\mu$ and the wavefunction in units of $\sqrt{\mu/g}$, where $\mu$ is the chemical potential of the condensate. Applying these scalings, the GPE is reduced to the dimensionless form:
\begin{equation}\label{eq:gpe}
    i\partial_t\psi(\vb{r}) = -\frac{1}{2}\vb{\nabla}^2\psi(\vb{r}) + |\psi(\vb{r})|^2\psi(\vb{r}).
\end{equation}
In this work, all simulations are based on the dimensionless form of the GPE described in Eq.~(\ref{eq:gpe}). The wavefunction can be expressed in terms of its density and phase using the Madelung transformation i.e. $\psi(\vb{r}) = \sqrt{\rho(\vb{r})}e^{i\theta(\vb{r})}$ where $\rho(\vb{r}) = |\psi(\vb{r})|^2$ is the density and $\theta(\vb{r})$ is the phase of the wavefunction. Using this transformation, the particle number current density corresponding to the wavefunction,
\begin{equation}    
\vb{J}(\vb{r}) = \frac{i}{2}\left[ \psi(\vb{r}) \vb{\nabla} \psi^*(\vb{r}) -  \psi^*(\vb{r}) \vb{\nabla} \psi(\vb{r}) \right]
\end{equation}
can be written in the form, $\vb{J}(\vb{r}) = \rho(\vb{r})\vb{v}(\vb{r})$.
\parshape=0
This defines the velocity field for the system as
\begin{equation}
\vb{v}(\vb{r}) = \vb{\nabla} \theta(\vb{r}).
\end{equation}
The velocity field is irrotational everywhere except at singularities in the wavefunction.

The total energy of system in the GPE is known to be conserved~\cite{Pethick:CUP2008, Pitaevskii:OUP2016}. The corresponding Hamiltonian for the GPE is given as
\begin{equation}\label{eq:hamiltonian}
    H = \int \left[ \frac{1}{2}|\vb{\nabla}\psi|^2 + \frac{1}{2}|\psi|^4\right] d\vb{r}.
\end{equation}
The first term in the Hamiltonian pertains to the kinetic energy ($E_{kin}$) of the system, which can further be decomposed into quantum ($E_{kin}^q$), incompressible ($E_{kin}^i$) and compressible kinetic energy ($E_{kin}^c$)~\cite{Bradley:PRA2022}. With this separation, the kinetic energy takes the following form
\begin{align}
     E_{kin} &= \int \frac{1}{2}|\vb{\nabla}\psi|^2 d\vb{r}\\
             &= \int \frac{1}{2}\left[ |\vb{\nabla} \sqrt{\rho}|^2  + |\vb{w}^i|^2 + |\vb{w}^c|^2 \right]d\vb{r}\\
             &= E_{kin}^q + E_{kin}^i + E_{kin}^c,
\end{align}
where $\vb{w}^i $ and $\vb{w}^c$ represent the incompressible and compressible component of the density weighted velocity field $\vb{w} = \sqrt{\rho}\vb{v}$. These components are a consequence of the Helmholtz decomposition of a vector field into a divergence-free (incompressible) component and a curl-free (compressible) component. The second term in Eq. (\ref{eq:hamiltonian}) corresponds to the interaction energy of the system arising from the nonlinear interactions within the atoms of the condensate and is given by
\begin{equation}
    E_{int} = \int \frac{1}{2}|\psi|^4 d\vb{r}.
\end{equation}
\subsection{Kinetic Energy Spectra}
The kinetic energy spectra at time $t$ is defined as
\begin{equation}
    E^{\{i,c,q\}}_{kin}(t) = \frac{1}{2}\int|\vb{w}^{\{i,c,q\}}(\vb{k})|^2d\vb{k} = \int \varepsilon^{\{i,c,q\}}_{kin}(k)dk,
\end{equation}
where $\vb{w}^q = \vb{\nabla} \sqrt{\rho}$. Following the methods outlined in ~\cite{Bradley:PRA2022}, we calculate high resolution spectra corresponding to incompressible, compressible and quantum kinetic energy using
\begin{equation}\label{eqn:keSpectrum}
    \varepsilon^{i,c,q}_{kin}(k) =  \frac{1}{2} \int d\vb{r} \, \Lambda(k, |\vb{r}|) \, C[\vb{w}^{i,c,q}, \vb{w}^{i,c,q}](\vb{r}),
\end{equation}
where $\Lambda(k, |\vb{r}|) = \frac{1}{2\pi} k J_0(kr)$, $J_0$ being the zeroth order Bessel function of the first kind and $C\left[\vb{a}, \vb{b}\right](r)$ is the two-point correlation between fields $\vb{a}$ and $\vb{b}$.
\begin{figure*}[!htp]
\centering
\includegraphics[width=\textwidth]{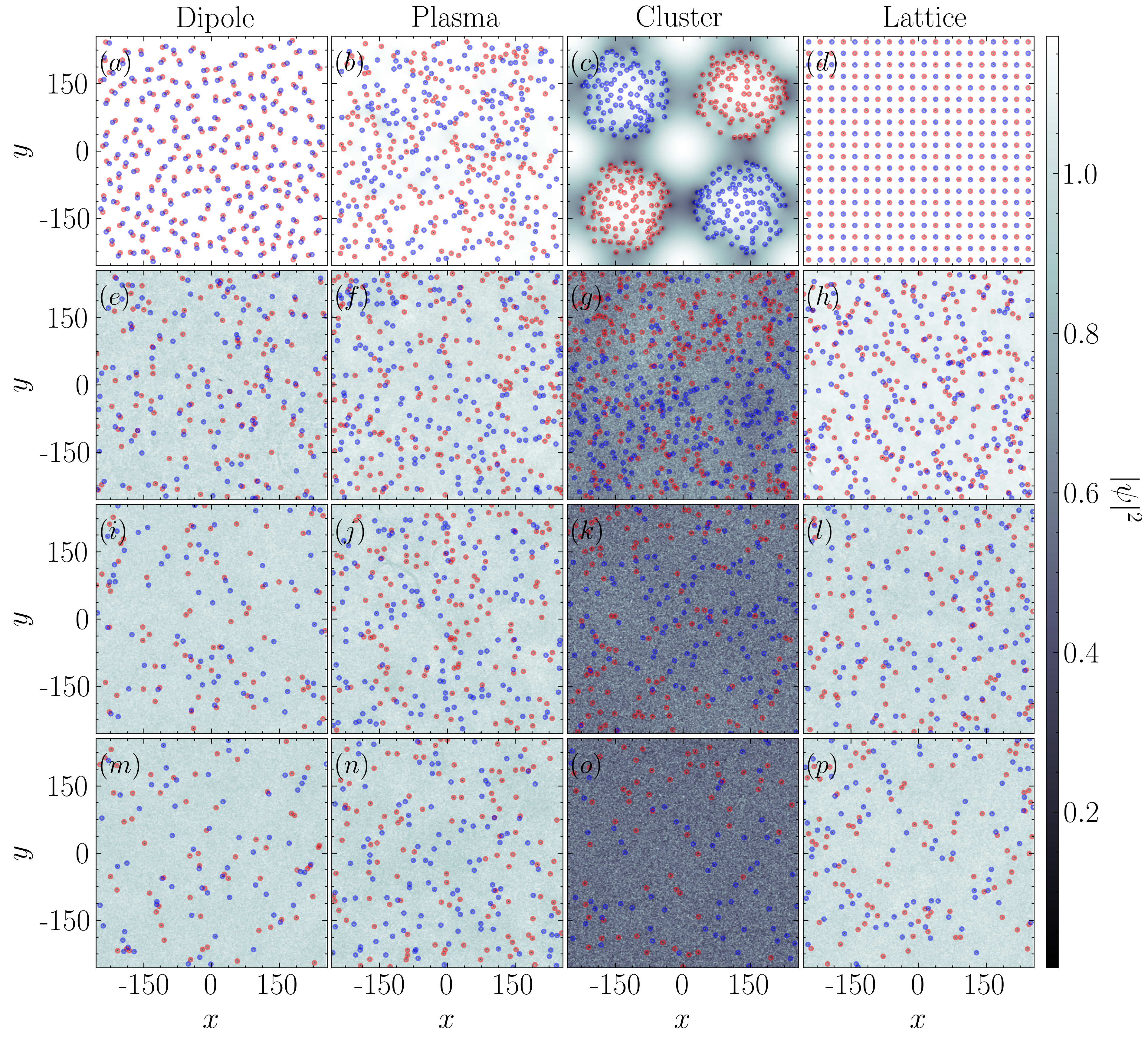}
\caption{The density of the wavefunctions for four different vortex configurations: (a) Dipole gas (b) Plasma (c) Cluster (d) Lattice. The blue markers are placed around vortices and the red markers around antivortices. Subsequent rows show the density of the wavefunction at times $t=0$, $10000$, $50000$ and $100000$ for the four distributions respectively.}
\label{fig:density}
\end{figure*}

\subsection{Particle number spectra and transfer function}
The total particle number ($N = \int |\psi|^2 d\vb{r}$) in the GPE is conserved. When the system is evolved in time, the system can only redistribute particle number across scales while keeping the total $N$ constant. To characterize how the particle number is distributed across different wavenumbers, we define the particle number spectra at time $t$ as
\begin{equation}
    N(t) = \int|\phi(\vb{k},t)|^2d\vb{k} = \int N(k,t)dk.
\end{equation}
where $\phi(\vb{k}, t)$ is the Fourier transform of $\psi(\vb{r},t)$. In practice, we perform angular binning to compute the particle number associated with different wavenumber shells. To analyze how the particle number is redistributed across scales, we next define the particle number transfer function, which measures the instantaneous rate of change in particle number for each mode. The Fourier transform of Eq. (\ref{eq:gpe}) gives
\begin{equation}\label{eq:gpek}
    i\partial_t\phi(\mathbf{k},t) = \frac{1}{2}k^2\phi(\mathbf{k},t) + R(\mathbf{k},t)
    \parshape=0.
\end{equation}
Here, $R(k)$ is the Fourier transform of the nonlinear term in the GPE (i.e. $R(\mathbf{k},t) = \mathcal{F} \left[ g|\psi(t)|^2\psi(t) \right]$). Multiplying $\phi^*(k,t)$ on both sides of the above equation yields
\begin{equation}
    i\phi^*(\mathbf{k},t)\partial_t\phi(\mathbf{k},t) = \frac{1}{2}k^2|\phi(\mathbf{k},t)|^2 + \phi^*(\mathbf{k}, t)R(\mathbf{k}, t)
    \parshape=0.\label{eq:Nk}
\end{equation}
Subtracting from this equation (Eq.~(\ref{eq:Nk})) it's complex conjugate yields the equation corresponding to the transfer rate given by
\begin{equation}
    \frac{\partial}{\partial t} N(\mathbf{k},t) = T_N(\mathbf{k},t) = 2\Im{\left[\phi^*(\mathbf{k}, t)R(\mathbf{k}, t)\right]}, 
    \parshape=0
    \label{eq:transfer}
\end{equation}
where $N(\mathbf{k}, t) = |\phi(\mathbf{k}, t)|^2$ is defined as the particle number in k-space and $T_N(\mathbf{k}, t) = 2\Im{\left[\phi^*(\mathbf{k},t)R(\mathbf{k},t)\right]}$ is the particle number transfer function. Here, $\Im[.]$ denotes the imaginary part of the function. 
\section{Numerical Results}
\label{sec:numresults}
In the present work, we consider four initial states that differ in their vortex-antivortex configurations, all defined within a periodic box of size $L = 512$. The first initial condition consists of a dipole gas where 200 pairs of vortices and antivortices, separated from each other by a distance of $3\pi$ (with a small uniform random perturbation added to the pair separation), are randomly distributed throughout the domain. The second is the plasma case where 200 vortices and 200 antivortices are each distributed randomly throughout the system. Next is an initial condition with four clusters, two of vortices and two of antivortices. Each cluster has 100 vortices of the same type randomly placed inside a circle of radius $L/8$. The center of the clusters are separated by a  distance of {$L/2$}. The fourth and final initial condition is composed of a lattice of 200 vortices and 200 antivortices such that each lattice site has vortices of the opposite charge as its nearest neighbor.

To generate the initial wavefunction for a particular configuration of vortices, we first compute the associated phase profile using the methods elaborated in Ref.~\cite{Billam:PRL2014}. Thereafter, starting with a uniform density, we evolve the wavefunction in imaginary time ($t\rightarrow -i t$)  while keeping the phase profile fixed. This procedure yields a wavefunction that faithfully contains the desired vortex configuration. Once the initial wavefunction is obtained for a particular vortex configuration, we evolve it in real time. For both imaginary and real time evolution of the condensate, we use the time-splitting spectral scheme (TSSP)~\cite{Rawat:CPC2025} and perform the simulation on a grid size of $(N_x, N_y) = (1024, 1024)$ with time step $dt = 0.001$.

\subsection{Dynamical evolution of different configuration of the vortices}
In the study of turbulence within BECs, the configuration and interaction of quantized vortices are central to the system's approach toward thermalization. The nonlinear dynamics of vortex structures dictate the complex evolution from non-equilibrium initial states to distinct thermalized regimes. Various initial vortex configurations, such as vortex lattices, clusters, or random vortex distributions, can significantly influence the routes and timescales of relaxation through mechanisms including vortex reconnections, annihilations, and sound emission~\cite{Barenghi:PNAS2014, White:PRA2012}. These processes mediate the transfer of energy and momentum within the condensate, guiding the turbulent system toward different thermalized states characterized by unique spectral and statistical properties~\cite{Bradley:PRX2012, Berloff:PRA2002}. In this section, we analyze the dynamical evolution of such vortex configurations in turbulent BECs to elucidate the underlying physical principles that govern quantum turbulence and thermalization dynamics.
\begin{figure}[!htp]
\centering
\includegraphics[width=1.0\linewidth]{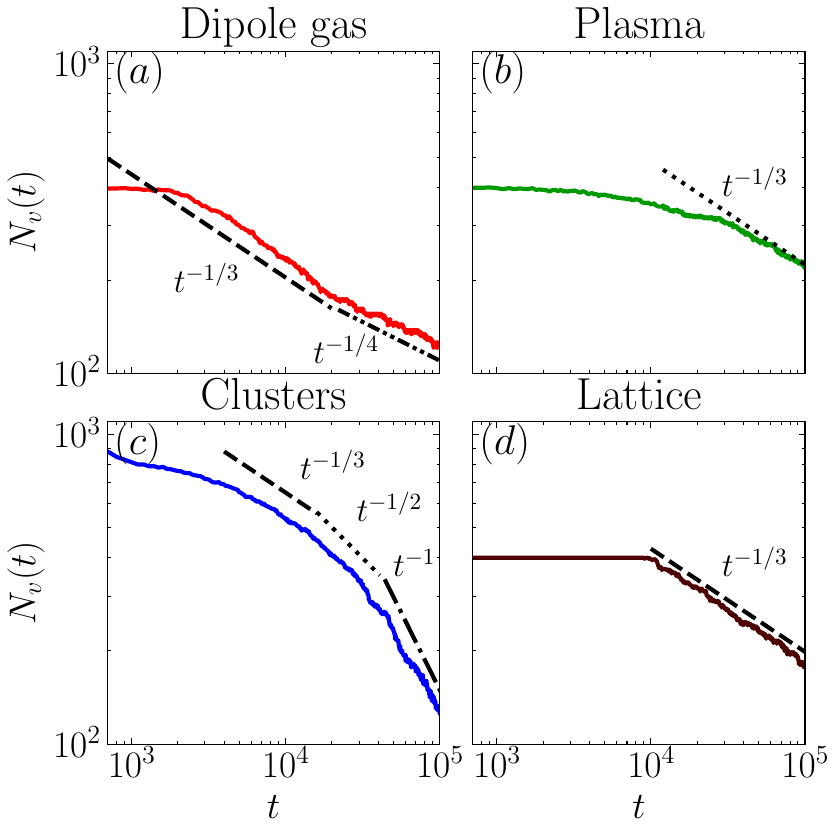}
\caption{The time evolution of the total number of vortices $N_{v}(t)$ present in the system  for the four different initial vortex-antivortex configurations: (a) Dipole, (b) Plasma, (c) Cluster, and (d) Lattice. In the dipole configuration, the vortex number decays with a scaling exponent of $-1/3$, followed by a slower decay with the smaller exponent of $-1/4$ at later times. The Plasma configuration undergoes a relatively steady decay with a scaling exponent $-1/3$. The cluster configuration undergoes a multi-stage decay, where the exponents decay from  $-1/3$ to $-1/2$, and eventually approaches  $-1$ at long times. In the lattice case, once the lattice structure melts, the subsequent decay of vortex number follows a $-1/3$ scaling, similar to the dipole case.}\label{fig:nv_time_evo}
\end{figure}

In Fig.~\ref{fig:density}, we present the pseudo-color density plots corresponding to the four distinct initial vortex configurations and their subsequent time evolution. Each column displays a configuration, while the rows represent the density of the condensate at four different times: $t=0$, $10^4$, $5\times 10^4$, $10^5$ (in dimensionless units). Vortices and antivortices are marked by blue and red dots, respectively. The first column (a,e,i,m) shows the dipole vortex configuration, where 200 vortex-antivortex pairs are initially imprinted in a regular dipolar pattern. The second column illustrates the plasma configuration, characterized by a random spatial distribution of 200 vortices and antivortices. The third column corresponds to the cluster configuration, where four compact clusters (two of vortices and two of antivortices) are imprinted, each containing 50 same-sign vortices. The fourth column represents the lattice configuration, consisting of a regular lattice of 200 vortex-antivortex pairs.
\begin{figure}[!htp]
\centering
\includegraphics[width=\linewidth]{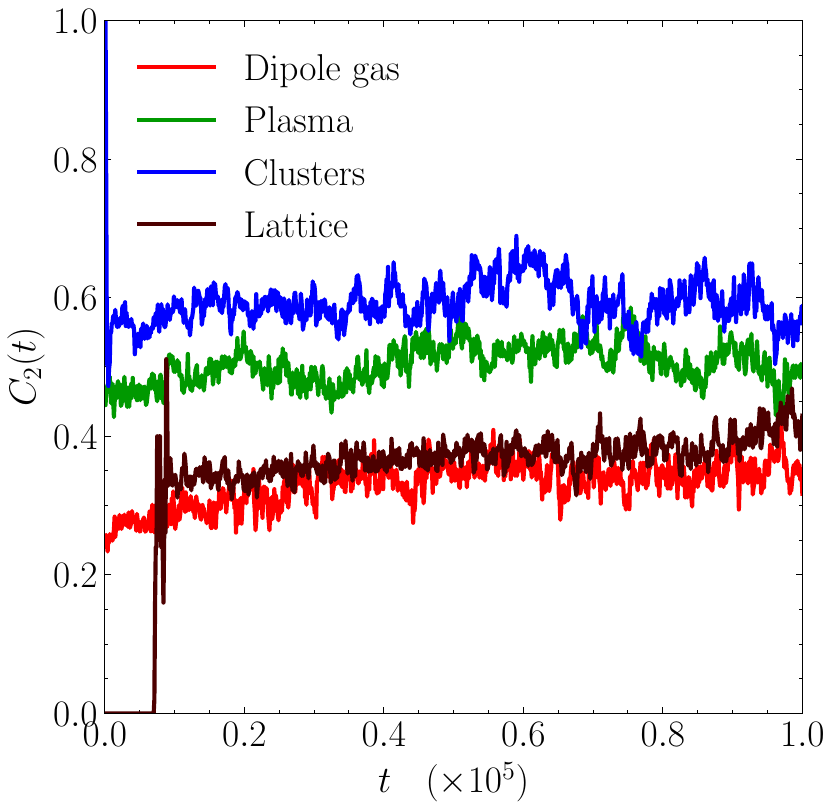}
\caption{Time evolution of the second-order vortex sign correlation function \( C_2 \) for four different vortex configurations (see legend for configuration types). The Dipole and Lattice configurations converge to a similar vortex configuration with \( C_2 \sim 0.38 \), while the Plasma configuration settles to \( C_2 \sim 0.5 \). The Cluster configuration stabilizes at \( C_2 \gtrsim 0.5 \), indicating different vortex dynamics for each configuration.}\label{fig:c2_time_evo}
\end{figure}

In the dipole configuration, a gradual reduction in vortex number is observed over time due to vortex-antivortex annihilations. The background color of the density plots transitions from white to light grey, indicating the generation of sound waves caused by pair-annihilation events. A similar trend is seen in the plasma configuration, although a larger number of vortices persist up to $t=10^5$, suggesting slower annihilation dynamics compared to the dipole gas.

In contrast, the cluster configuration displays qualitatively different behavior. The initially imprinted clusters expand, interact, and eventually merge, leading to strong nonlinear dynamics and significant sound wave emission, evident from the darker gray background. Despite the turbulence, by $t=10^5$, a striking spatial separation is observed: most vortices accumulate in the upper half, while most antivortices reside in the lower half of the condensate.

In the lattice configuration, the initially ordered vortex-antivortex structure begins to break down as early as $t=10^4$, with vortex-antivortex annihilations dominating the subsequent evolution. This leads to a rapid decrease in the total number of vortices. By $t=10^5$, the system has lost all of its initial order, and the resulting vortex distribution and background sound field closely resemble those observed in the dipole configuration at the same time.

To understand the dynamical behavior and decay mechanisms of different vortex configurations in two-dimensional (2D) quantum turbulence, it is essential to analyze the time evolution of the total number of vortices in the system. This quantity provides a direct measure of vortex-antivortex annihilation, which governs the decay of turbulent energy and enstrophy in superfluid systems \cite{Neely:PRL2013, Kwon:PRA2014, Stagg:PRA2015, Cidrim:PRA2016, Baggaley:PRA2018, Groszek:PRA2018, Kanai:PRA2025}.
At absolute zero, an isolated vortex-antivortex pair moves at a constant velocity while maintaining a fixed separation. Consequently, the observed decay of vortices in a turbulent ensemble necessitates interactions involving additional vortices or sound. Previous studies have identified two principal classes of annihilation mechanisms, a four-body mechanism and a three-body mechanism, distinguished by the number of vortices participating in the interaction~\cite{Cidrim:PRA2016, Baggaley:PRA2018, Groszek:PRA2018}. In the four-body mechanism, two vortices first form a bound state through interaction with a third vortex~\cite{Groszek:PRA2018}. The bound state subsequently decays to sound waves through interactions with another vortex. In the three-body mechanism, two vortices annihilate by approaching each other through interaction with a third vortex and the sound waves present in the system.
These decay of vortex number obey a rate equation given by $N_v\sim t^{-1/(n-1)}$~\cite{Baggaley:PRA2018, Kanai:PRA2025} where $n$ is the number of vortices participating in the annihilation process.
While these mechanisms have been primarily characterized in the context of randomly distributed vortex ensembles, we aim to investigate the extent to which the decay dynamics depends on the structure and topology of the initial state. By comparing the decay rate across different initial configurations such as dipole gases, vortex plasmas, clustered states, and lattices, we can probe how topology and coherence influence the relaxation pathways and overall decay of two-dimensional quantum turbulence.

In the dipole gas case [Fig.~\ref{fig:nv_time_evo}(a)], the number of vortices exhibits an initial power-law decay approximately following $ ( t^{-1/3} )$, which can be attributed to the four-body annihilation processes, as observed in previous studies. At later times, the decay slows down, with the effective scaling approaching $( t^{-1/4} )$. A likely factor responsible for this transition in the scaling exponent is the velocity dependence of vortices on vortex density~\cite{Karl:NJP2017}.
The plasma case [Fig.~\ref{fig:nv_time_evo}(b)] shows a very slow initial decay, approximately constant, which persists up to $( t \approx 3 \times 10^4 )$, before transitioning to a steeper scaling close to the four-body scaling $( t^{-1/3} )$. Similar early-time slow decay rates have been reported in previous studies~\cite{Groszek:PRA2016, Kanai:PRA2025}. This behavior reflects a suppressed annihilation rate due to the initial spatial randomness and low vortex-antivortex overlap.

The cluster configuration [Fig.~\ref{fig:nv_time_evo}(c)] exhibits the most distinct dynamics. Due to strong velocity gradients in the inter-cluster regions, a significant number of new vortex-antivortex pairs are generated in the early stages, a phenomenon consistent with prior observations of turbulent cluster interactions~\cite{Groszek:SPP2020}. The vortex number initially decays as $( t^{-1/3} )$, transitions to $( t^{-1/2} )$ around $( t = 2 \times 10^4 )$, and ultimately reaches to a rapid decay regime with scaling close to $( t^{-1} )$ by the end of the simulation. A similar behavior has been reported previously in finite temperature GPE simulations~\cite{Baggaley:PRA2018}. This correspondence suggests that the intense amount of sound waves present in the system plays a role analogous to that of a thermal component~\cite{Nazarenko:JLTP2007, Kanai:PRA2025}, enabling vortex dipoles to form, lose energy and decay without requiring interactions with other vortices. 
In the lattice case [Fig.~\ref{fig:nv_time_evo}(d)], the vortex number remains approximately constant as long as the ordered structure persists. Once the lattice destabilizes due to dynamical instabilities, the decay process starts and the vortex number transitions to a  $(t^{-1/3})$  scaling, similar to that observed during the initial stage of the dipole gas evolution, indicating a clear dominance of four-body annihilations in this regime.

To gain deeper insight into the evolving vortex dynamics arising from different initial configurations, we quantify the degree of order in the system using the second-order vortex sign correlation function $C_2$, defined as
\begin{equation}
C_2 = \frac{1}{2N_v}\sum_{i=1}^{N_v}\sum_{j=1}^2 c_{ij}, 
\end{equation}
where $N_v$ is the total number of vortices, and $c_{ij} = 1$ if the $i$\textsuperscript{th} vortex and its $j$\textsuperscript{th} nearest vortex have the same circulation sign (i.e., both are vortices or both are antivortices); otherwise $c_{ij} = 0$~\cite{Jha:PRF2025}. The value of $C_2$ serves as a diagnostic for the type of vortex configuration present in the system. Specifically, $C_2>0.5$ indicates predominance of like-signed vortex clusters, $C_2<0.5$ reflects a dipole-dominated regime, and $C_2=0.5$ corresponds to a random configuration of the vortices, with roughly equal contributions from clusters, dipoles, and free vortices.

We present the time evolution of the second-order vortex sign correlation function $C_2$, for various initial vortex configurations in Fig.~\ref{fig:c2_time_evo}. For the dipole case (red line), $C_2$ begins near $0.25$ and gradually increases to  $0.35$, where it stabilizes. Throughout the evolution $C_2$ remains below $0.5$, clearly indicating that vortex-antivortex dipoles dominate the system's dynamics. In contrast, the plasma case (green line) maintains a nearly constant $C_2\sim 0.5$ suggesting a random distribution of vortex signs that persists over time. For the cluster case (blue line), $C_2$ decreases rapidly from an initial value of $1$ to about $0.6$, due to the creation of new vortex-antivortex pairs as discussed earlier, and subsequently fluctuates around this level. The sustained value above $0.5$ indicates the continued dominance of vortex clustering. Interestingly, the lattice case follows a trajectory reminiscent of the dipole scenario. After the breakdown of the initial lattice structure, the system quickly transitions to $C_2 \approx 0.34$, which gradually increases and settles around $0.42$. This steady value below $0.5$ reflects a dipole-dominated regime, albeit with a slightly stronger tendency toward clustering compared to the pure dipole case.

\begin{figure}
\centering
\includegraphics[width=\linewidth]{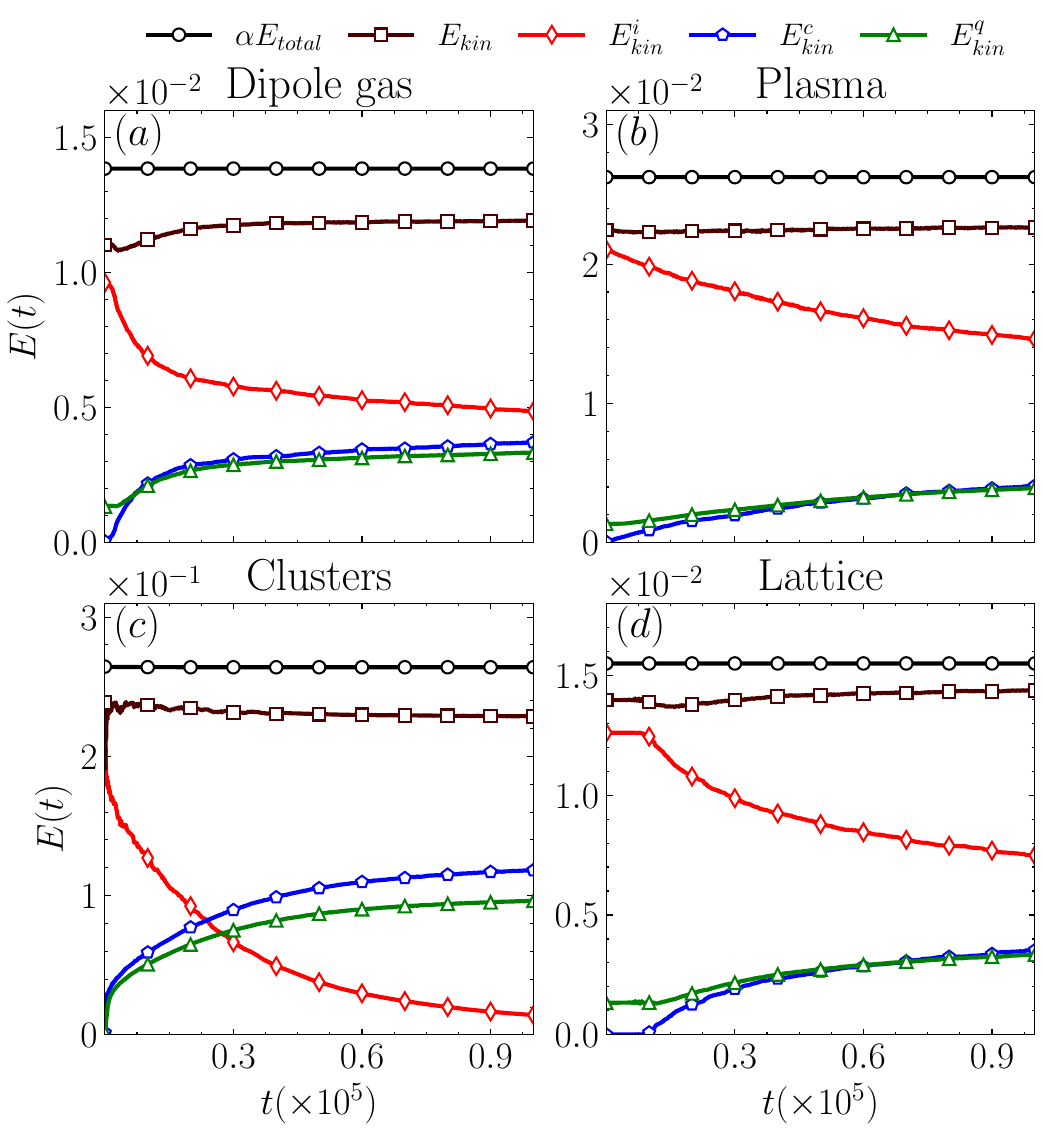}
\caption{The time evolution of kinetic energy and its components for the four different vortex configurations: (a) dipole, (b) plasma, (c) cluster, and (d) lattice. The blue line represents the total kinetic energy (\(E_{kin}\)), the red line represents incompressible kinetic energy (\(E^{i}_{kin}\)), the green line represents compressible kinetic energy (\(E^{c}_{kin}\)), and the brown line represents quantum kinetic energy (\(E^{q}_{kin}\)). The black line represents the total energy of the system, rescaled by a factor \(\alpha\), with values of 0.03 and 0.09 for the top and bottom rows, respectively. The depletion in incompressible kinetic energy reflects the vortex-antivortex annihilations. The energy extracted from the annihilation processes are transferred to other components: compressible KE, quantum KE and interaction energy. Differences in the rate and magnitude of this redistribution across the four cases arise from the distinct initial vortex configurations, which govern how efficiently dipoles form, interact, and annihilate.}\label{fig:ke_time_evo}
\end{figure}

To further characterize the system's dynamics, in Fig.~\ref{fig:ke_time_evo}, we examine the temporal evolution of the different energy components for the various initial vortex configurations. The total energy (black lines with circular markers) has been rescaled in each case to enhance the visibility of the individual energy contributions. The total kinetic energy of the system is shown in brown with square markers, while the incompressible, compressible, and quantum kinetic energies are represented by red (diamonds), blue (pentagons), and green (triangles), respectively.

In all the cases, the total energy remains constant with time as expected. The observed decay in the incompressible kinetic energy, \( E_{kin}^i \), reflects a redistribution of energy into the compressible and quantum kinetic channels. This transfer is primarily driven by vortex–vortex and vortex–sound interactions~\cite{Nore:POF1997, Numasato:PRA2010}, both of which facilitate vortex–antivortex annihilation events that subsequently emit sound into the system. While these mechanisms are common to all cases, we demonstrate that their efficiency and characteristic timescales are strongly influenced by the initial vortex configuration.

In the dipole case as illustrated in the Fig.~\ref{fig:ke_time_evo}(a), the total incompressible kinetic energy starts at the value of $E_{kin}^i(0)=9.6\times10^{-3}$ and undergoes a rapid decay during the early stages of evolution. This rapid decay can be attributed to the annihilation of vortices, which typically requires the formation of vortex dipoles first. These pairs then interact with other vortices, as well as with the surrounding sound field, facilitating their annihilation. Notably, in this case, a significant number of dipoles are already present at the outset of the simulation, which allows the annihilation process to proceed immediately after the system is initialized. As the system evolves, the rate of decay slows down. This deceleration in the decay process is due to the increasing disruption of vortex pairs as they interact with one another and with other components of the system. Some of these vortex pairs become isolated, and the interactions between them decrease, leading to a reduced frequency of annihilation events over time. This two-stage behavior is also evident in the vortex number evolution [see Fig.~\ref{fig:nv_time_evo}], which shows an initial decay scaling as $ t^{-1/3}$, transitioning to a slower decay with a scaling as $ t^{-1/4} $ at later times.

In the plasma configuration [Fig.~\ref{fig:ke_time_evo}(b)], the system starts with a higher incompressible kinetic energy of $E_{kin}^i(0) = 2.10 \times 10^{-2}$, which decays more gradually, reaching $( 1.46 \times 10^{-2} )$ by the end of the simulation. Here, the vortices are initially distributed in a disordered manner, with many remaining unpaired. This delays annihilation, as oppositely signed vortices must migrate into proximity. Consequently, the decay of $ E_{kin}^i$ is slower compared to the dipole case. The corresponding vortex number also decreases much more gradually; see Fig.~\ref{fig:nv_time_evo}.

In the cluster configuration [Fig.~\ref{fig:ke_time_evo}(c)], the incompressible kinetic energy is the largest among all configurations, starting at $E_{kin}^i = 2.4 \times 10^{-1}$. This high initial energy arises from the strong velocity fields induced by clustered vortices and the spontaneous generation of additional vortex–antivortex pairs between clusters at early times. While many of these newly formed pairs are short-lived and quickly annihilate, they leave behind significant compressible sound energy. As a result, this configuration exhibits the highest compressible kinetic energy toward the end of the simulation. 

By contrast, the lattice configuration [Fig.~\ref{fig:ke_time_evo}(d)] behaves similarly to the dipole and plasma cases once the initial structure breaks down. The incompressible kinetic energy decays from $1.3 \times 10^{-2}$ to $7.5 \times 10^{-3}$ over time. Notably, among all four configurations, only in the cluster case does the compressible kinetic energy surpass the incompressible kinetic energy during the latter stages of evolution. Having examined the vortex annihilation dynamics and their role in guiding the system toward its final state across different initial configurations, we now turn our attention to the distribution and evolution of the velocity and particle number spectra in Fourier space.

\begin{figure}
\centering
\includegraphics[width=\linewidth]{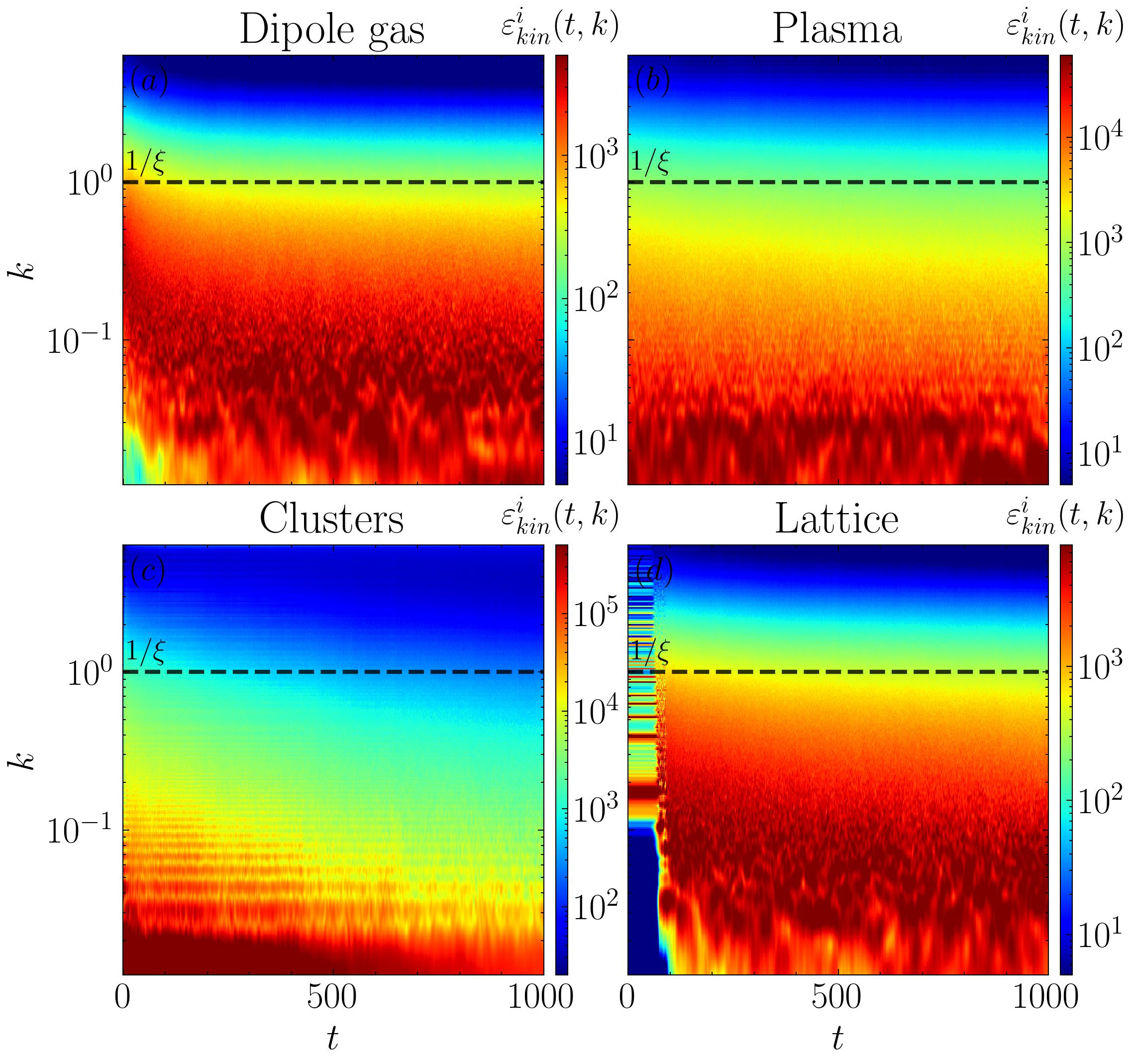}
\caption{A heatmap showing the time evolution of the incompressible kinetic energy ($\varepsilon^i_{kin}$) spectra for four different vortex configurations: (a) dipole, (b) plasma, (c) cluster, and (d) lattice. In all cases, the spectral intensity shifts toward lower wavenumbers over time, indicating an inverse cascade. The cluster case shows additional depletion at low wavenumbers and energy transfer from high $k$ modes to intermediate scales, due to the dynamic instability of the initially perfect clusters.}\label{fig:Eki_heatmap}
\end{figure}
\subsection{Effect of initial vortex configuration on the Kinetic energy spectra }
Understanding the distribution of energy across different length scales is fundamental to characterizing the nature of turbulence in BECs. Spectral analysis of the condensate offers key insights into the mechanisms governing relaxation and thermalization, including energy cascades, and dissipative processes~\cite{Bradley:PRX2012, Bradley:PRA2022}. In this section, we investigate the incommpressible and compressible energy spectra, as defined in Eq.~(\ref{eqn:keSpectrum}), to elucidate the multiscale structure of the turbulent flow and assess its role in the subsequent relaxation dynamics of the system.

In Fig.~\ref{fig:Eki_heatmap}, we present heatmaps showing the time evolution of the incompressible kinetic energy spectra for the four distinct initial vortex configurations considered in our work. These heatmaps provide a visual representation of how energy is redistributed across spatial scales as the system evolves.  The horizontal axis corresponds to time, while the vertical axis represents the wavenumber $k$, with the color scale indicating the spectral energy density. A dashed horizontal line marks the inverse healing length $1/\xi$ which approximately separates the hydrodynamic regime $(k<1/\xi)$ from the vortex-core region $(k>1/\xi)$. 

In the dipole gas case [Fig.~\ref{fig:Eki_heatmap}(a)], we observe a clear shift of spectral energy density toward lower wavenumbers over time, evidenced by the movement of the spectral maxima (red region) to the lower $k$ region.  This reflects the annihilation of numerous dipoles and the disruption of others due to vortex-vortex interactions, leading to the generation of larger-scale flow structures. By contrast, the plasma case [Fig.~\ref{fig:Eki_heatmap}(b)] shows only a weak transfer of incompressible kinetic energy toward low wavenumbers, with noticeable changes occurring mainly at intermediate $k$ values. This suggests that, although small clusters may form, break down and reform dynamically, there is no net growth of clustering in this system. Vortices continually reorganize into transient multi-vortex structures, but these fluctuations do not accumulate into persistent large-scale organization, and consequently no substantial large-scale flow is generated. These spectral features therefore align with the behavior of the second-order vortex sign correlation discussed in Fig.~\ref{fig:c2_time_evo}.
\begin{figure}[!htp]
\centering
\includegraphics[width=\linewidth]{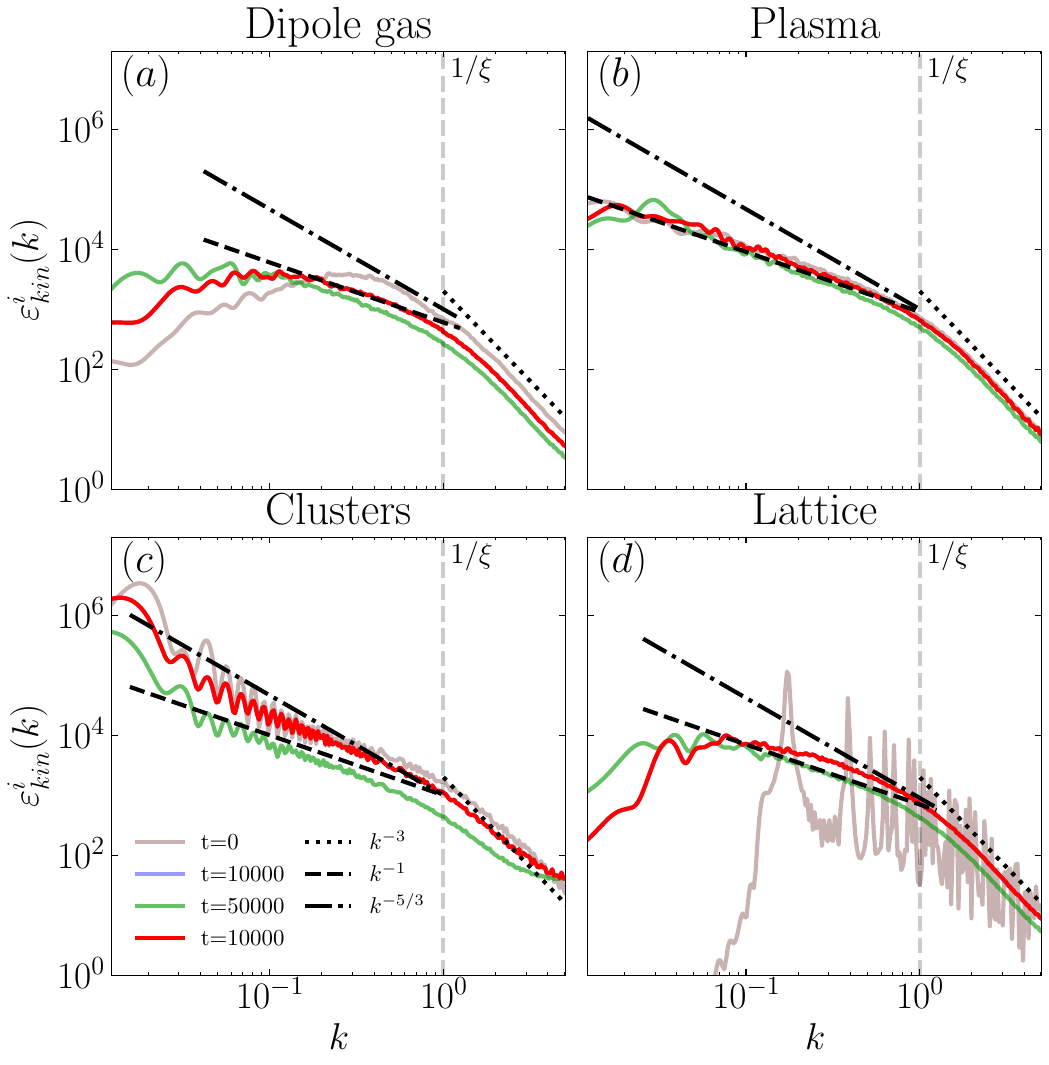}
\caption{The incompressible kinetic energy spectra ($\varepsilon^i_{kin}$) as function of the wavenumber for the four different vortex distributions: (a) Dipole, (b) Plasma, (c) Cluster, and (d) Lattice. For the dipole, plasma, and lattice initial configuration of the vortices, the energy spectra shows a Vinen-like scaling ($\varepsilon^{i}_{kin}\sim k^{-1}$) in the infrared regime. In contrast, the cluster configuration exhibits a Kolmogorov-like scaling ($\varepsilon^{i}_{kin}\sim k^{-5/3}$).  }\label{fig:Spectra_Eki}
\end{figure}
The cluster case [Fig.~\ref{fig:Eki_heatmap}(c)] demonstrates a distinct behavior. A significant amount of energy is initially concentrated at low wavenumbers, consistent with the presence of large, like-signed vortex clusters dominating the early-time dynamics. However, as time progresses, the clusters gradually lose coherence, in part due to the formation of dipoles between them. As a result, the spectral energy content diminishes across all scales. In the lattice case, once the regular vortex structure breaks down, the spectral dynamics resemble those of the dipole gas. A noticeable energy transfer from high to low wavenumbers occurs, reflecting the transition toward larger-scale, dipole-driven flow features [See Fig.~\ref{fig:Eki_heatmap}(d)].
\begin{figure}[!htp]
\centering
\includegraphics[width=\linewidth]{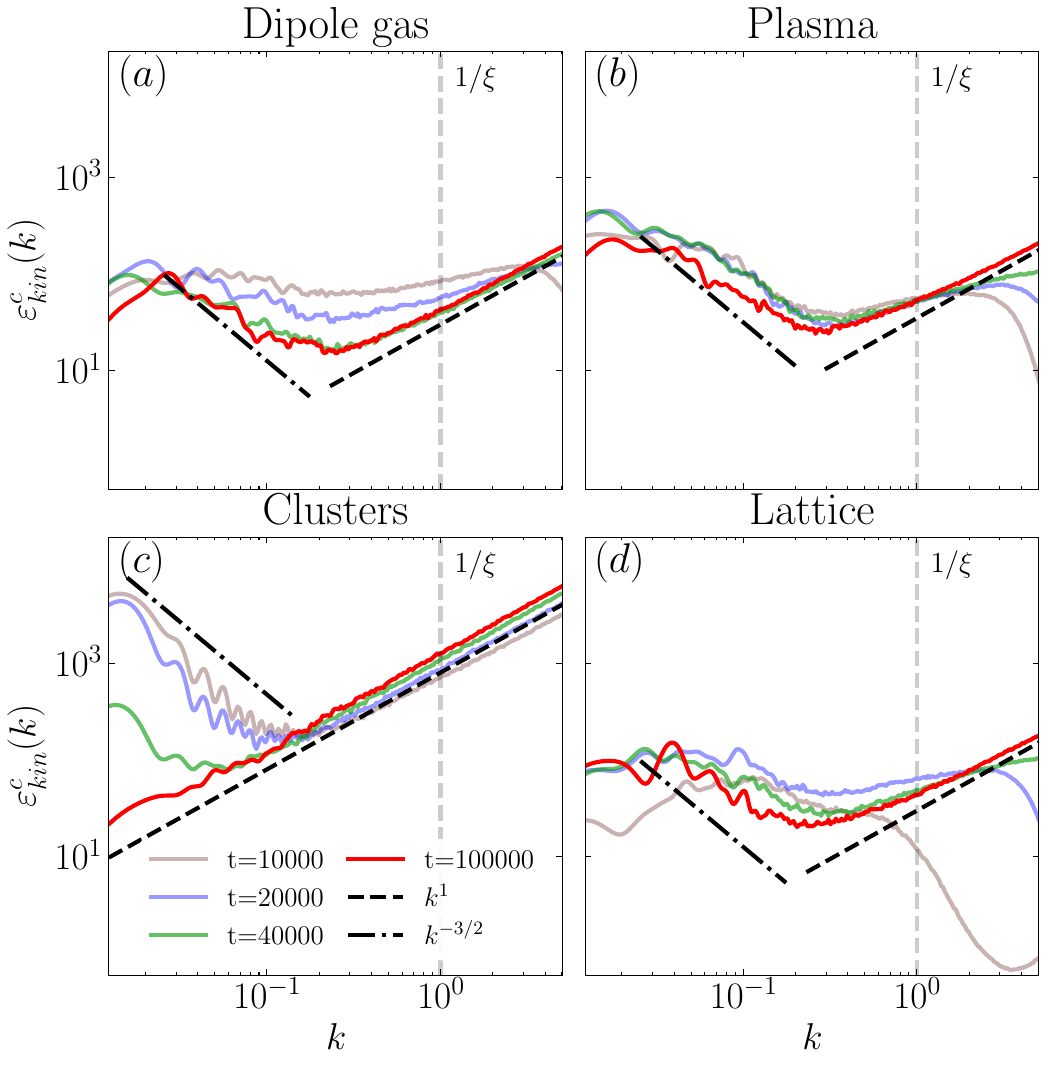}
\caption{
The compressible kinetic energy spectra ($\varepsilon^c_{kin}(k)$) as a function of the wavenumber for four different initial vortex distributions: (a) Dipole, (b) Plasma, (c) Cluster, and (d) Lattice. The dipole, plasma, and lattice configurations exhibit partial thermalization, characterized by a $k^{1}$ scaling in the ultraviolet region, while a $k^{-3/2}$ scaling, indicative of weak-wave turbulence, is observed in the infrared region. In contrast, the cluster configuration displays a fully thermalized spectrum across the entire accessible range of wavenumbers.}\label{fig:Spectra_Ekc}
\end{figure}

To further examine how the incompressible kinetic energy is distributed across spatial scales at specific stages of the evolution, we analyze the energy spectra at selected time instants in order to identify the scaling laws that emerge in each vortex configuration. In Fig.~\ref{fig:Spectra_Eki}, we show the incompressible kinetic energy spectra for the four vortex distributions under consideration. Irrespective of the initial configuration, the spectra exhibit a universal scaling of $k^{-3}$ in the ultraviolet regime ($k > 1/\xi$), characteristic of the vortex core structure~\cite{Bradley:PRX2012, Bradley:PRA2022}.

For the dipole gas [Fig.~\ref{fig:Spectra_Eki}(a)], a Kolmogorov-like $k^{-5/3}$ scaling emerges at a narrow band of infrared (IR) wavenumbers during early times. This is likely a consequence of a direct cascade regime dominated by the decay of vortex dipoles present in the system at the early stages of evolution~\cite{Numasato:JLTP2010, Numasato:PRA2010}. The scaling in the spectra gradually transitions to a $k^{-1}$ over a broad IR range at later times. The $k^{-1}$ scaling in the IR region of the incompressible kinetic energy spectrum is generally associated with a random distribution of vortices lacking dominant large-scale coherence~\cite{Bradley:PRX2012}. This transition reflects the progressive randomisation of the vortex distribution: the initially well-organized dipolar configuration becomes increasingly disordered as dipoles annihilate or are disrupted, leading to the appearance of free vortices and small clusters, while still remaining predominantly dipole-dominated. Accordingly, in the plasma case [Fig.~\ref{fig:Spectra_Eki}(b)], the spectra exhibits a persistent $k^{-1}$ scaling throughout the evolution, consistent with the corresponding $C_2$ values fluctuating around 0.5; see Fig.~\ref{fig:c2_time_evo}.

In the case of vortex clusters, a significant accumulation of incompressible energy is observed at large spatial scales, evidenced by a prominent hump in the small-$k$ range in Fig.~\ref{fig:Spectra_Eki}(c) during early times. The spectrum in this case shows a scaling close to $k^{-5/3}$, reflecting the influence of coherent, large-scale vortex structures. However, this scaling transitions gradually towards $k^{-1}$ driven by the loss of cluster coherence from the emergence of newly formed vortex dipoles, as discussed earlier. The $k^{-5/3}$ scaling re-emerges later likely due to an inverse cascade process driven by evaporative heating of dipoles~\cite{Simula:PRL2014}. Finally, in the lattice configuration, once the regular vortex arrangement breaks down, the spectrum evolves toward a $k^{-1}$ scaling, consistent with the increasing randomization of vortex positions and the formation of dipole-like structures [see Fig.~\ref{fig:Spectra_Eki}(d)].

In Fig.~\ref{fig:Spectra_Ekc}, we present the compressible kinetic energy spectra corresponding to various initial vortex configurations. For the dipole, plasma, and lattice cases, the spectra in the infrared (IR) regime exhibit an approximate power-law scaling  $k^{-3/2}$. This behavior is indicative of a weak wave turbulence cascade, characteristic of sound wave interactions in the condensate~\cite{Nazarenko:JLTP2006, Reeves:PRA2012}. At later times, these same configurations display a spectral scaling $k$ in the ultraviolet (UV) regime, suggesting the onset of partial thermalization of the compressible modes at high wavenumbers.
\begin{figure}
\centering
\includegraphics[width=\linewidth]{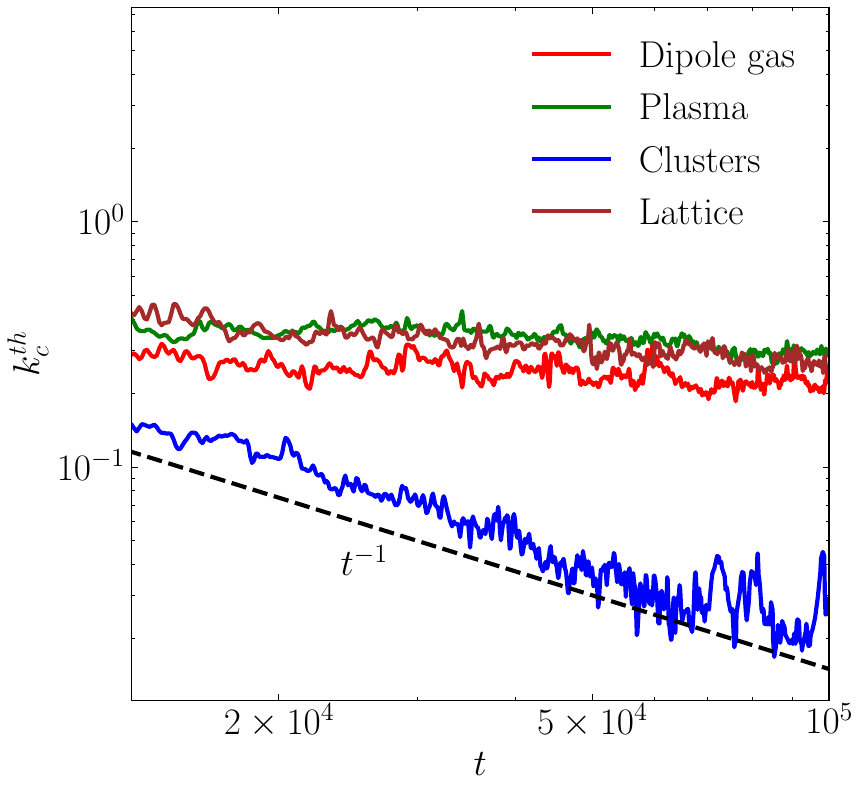}
\caption{The temporal evolution of the critical wavenumber ($k_c^{th}$), beyond which the compressible kinetic energy spectrum, shown in Fig.~\ref{fig:Spectra_Ekc}, exhibits a $k$-scaling characteristic of thermalization, for different vortex initial configurations. In the case of the cluster, $k_c^{th}$ rapidly shifts towards lower wavenumbers, following an approximate $t^{-1}$ scaling, whereas in the other cases, the drift is minimal.}\label{fig:bend_loc_Ekc}
\end{figure}
In contrast, the clustered vortex configuration shows a markedly different behavior. Its compressible kinetic energy spectrum follows a $k$ scaling across the entire wavenumber range at later times (specifically around $t \sim 10^5$). This observation is consistent with the complete thermalization of the compressible component, indicating that all available modes have reached thermal equilibrium\cite{Numasato:PRA2010}. On the other hand, for the dipole, plasma, and lattice cases, the $k$ scaling appears only beyond a certain wavenumber threshold, implying that thermalization is restricted to the high-$k$ modes, while the lower modes remain non-thermalized.

To gain further insight into the approach toward thermal equilibrium, we compute the critical wavenumber $k_c^{\text{th}}$, defined as the threshold beyond which the compressible energy spectrum follows the thermal $k$ scaling. The time evolution of $ k_c^{\text{th}}$ for each vortex configuration is shown in Fig.~\ref{fig:bend_loc_Ekc}. For the cluster case (depicted by the blue line), $k_c^{\text{th}}$ exhibits a clear power law decay with time with an exponent $-1$, signifying a complete thermalization of the system. In contrast, for the dipole, plasma, and lattice configurations, $k_c^{\text{th}}$ rapidly stabilizes and remains nearly constant over time. This plateau suggests that these systems become dynamically arrested in a non-thermal state, where only a limited range of high-$k$ modes attain thermal characteristics.

\subsection{Detection of thermalization using the energy transfer scheme}
\begin{figure}
\centering
\includegraphics[width=\linewidth]{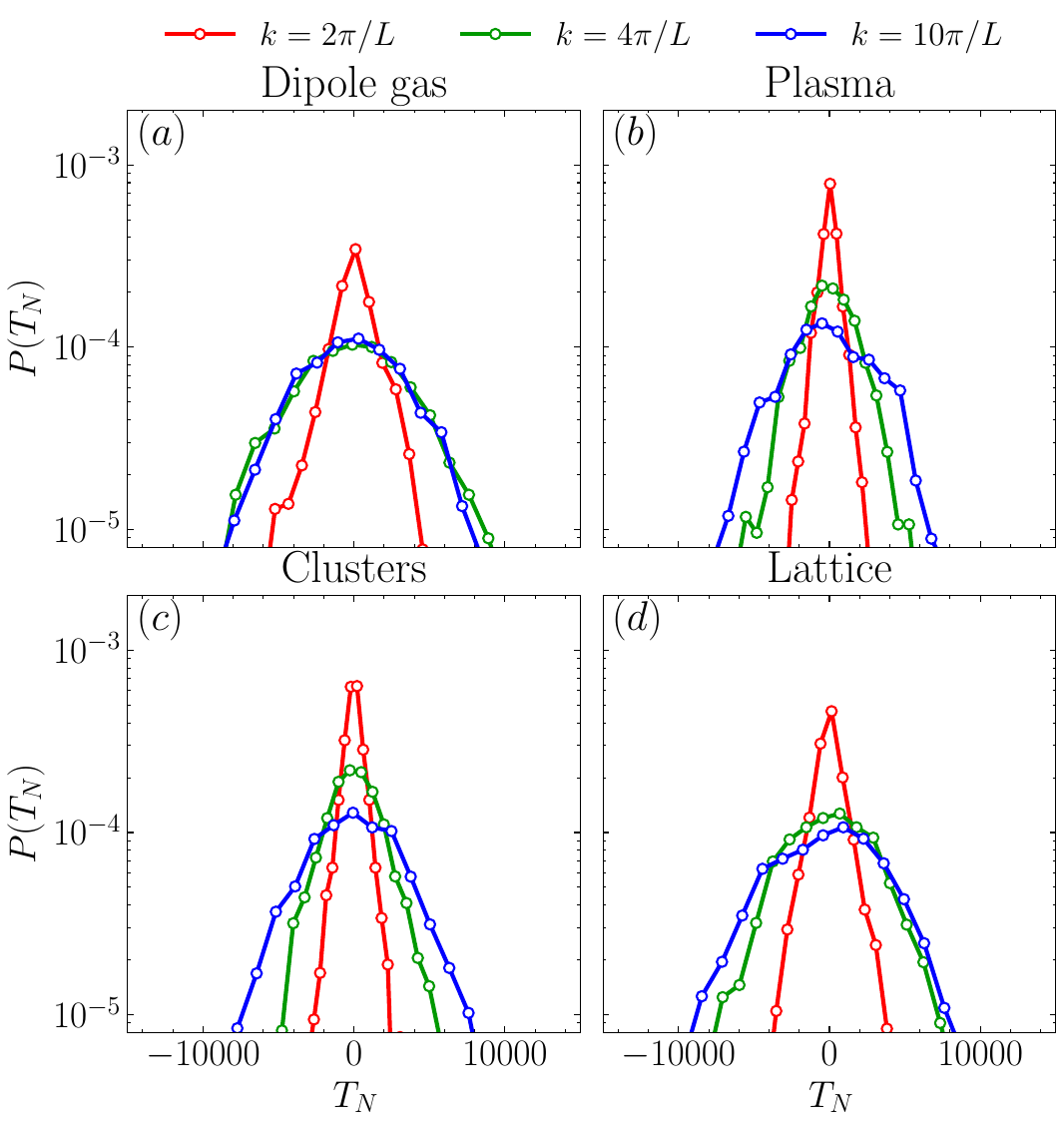}
\caption{PDF of $T_N(k,t)$ obtained for specific wavenumber shells $k = 2\pi/L$, $4\pi/L$ and $10\pi/L$ for the different vortex configurations: (a) Dipole, (b) Plasma, (c) Cluster and (d) Lattice. For the cluster case [(c)], the PDF shows a symmetric Gaussian distribution with zero mean, indicating a zero mean rate of energy transfer-ideal for equilibrium. In contrast, the PDFs for other configurations [(a), (b), (d)] are skewed, indicating non-equilibrium behavior.}\label{fig:Tk_pdf}
\end{figure}

BECs initialized with non-equilibrium configurations of vortices and anti-vortices provide a rich platform for studying thermalization in quantum fluids. The dynamics involve complex vortex motion, annihilation events, and energy cascades across spatial and temporal scales. Traditional probes of thermalization, such as monitoring vortex number decay or phase coherence, often fail to capture the underlying mechanisms of energy redistribution. To address this, we employ an energy transfer scheme to study the evolution of different modes in the condensate. In a system close to thermal equilibrium, energy exchange is expected to satisfy detailed balance and thermal fluctuation relations, while deviations indicate non-equilibrium phases, prethermalization plateaus, or incomplete relaxation.
\begin{figure}[!htp]
\centering
\includegraphics[width=\linewidth]{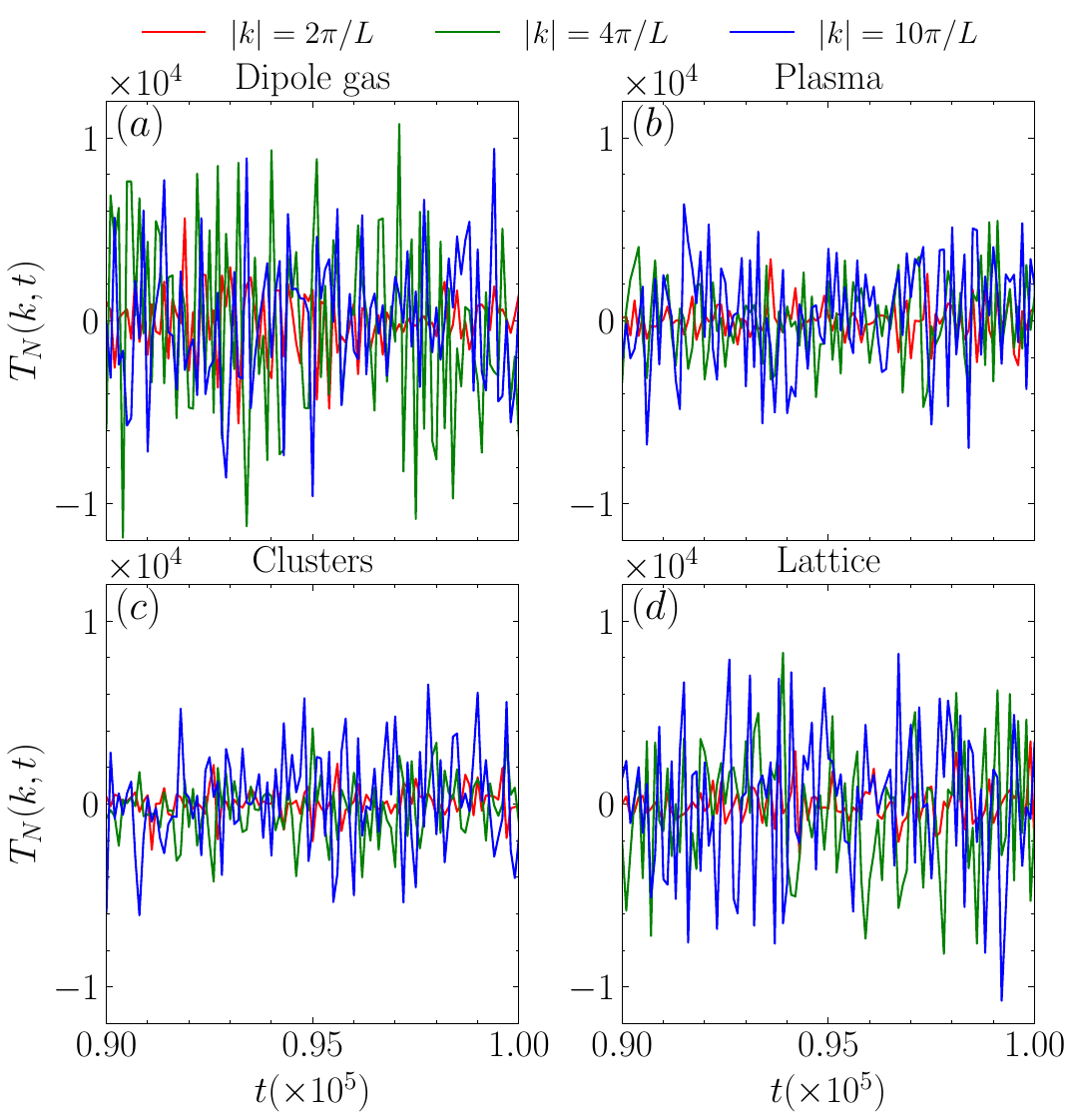}
\caption{The time evolution of the particle number transfer function $T_N(k,t)$ [as defined in Eq.~(\ref{eq:transfer})] in shells $k = 2\pi/L$ (blue), $4\pi/L$ (orange) and $10\pi/L$ (green) for the different initial vortex distributions: (a) Dipole, (b)Plasma, (c) Cluster and (d) Lattice. All modes exhibit irregular sign-changing fluctuations, reflecting the persistent and configuration-dependent redistribution of particle number across scales.}\label{fig:Tk}   
\end{figure}
In Fig.~\ref{fig:Tk}, we show the temporal evolution of the particle number transfer function \( T_N(k,t) \) [Eq.~(\ref{eq:transfer})] for the large-scale modes across the four cases considered in our study. We observe that \( T_N(k,t) \) exhibits significant fluctuations in the wavenumber shells \( |k| = 2\pi/l \), \( 4\pi/l \), and \( 10\pi/l \), even after long evolution times. To further investigate the erratic behavior of \( T_N(k,t) \) between \( t = 0.9 \times 10^5 \) and \( t = 1.0 \times 10^5 \), we analyze the probability density function (PDF) of the transfer function, \( P(T_N) \), as shown in Fig.~\ref{fig:Tk_pdf}. The PDF is computed from the values of \( T_N(k) \) obtained between \( t = 0.5 \times 10^5 \) and \( t = 1.8 \times 10^5 \) to eliminate the influence of any early-time transient dynamics arising from the different initial vortex configurations.

In the cluster case, we find that the PDF is Gaussian in nature, with a zero mean for all modes (\( |k| = 2\pi/L \), \( 4\pi/L \), \( 10\pi/L \)), indicating a zero mean rate of energy transfer, a characteristic typically associated with equilibrium states. In contrast, for the other initial conditions—dipole gas [(a)], plasma [(b)], and lattice [(c)]—the temporal PDFs of all large-scale modes exhibit skewed Gaussian distributions, indicating a non-zero mean and thus a non-equilibrium state for these configurations. This behavior highlights the distinctive nature of the energy transfer processes in these cases, where the distribution of energy transfer is asymmetric and indicative of persistent fluctuations away from equilibrium.

\begin{figure}
\centering
\includegraphics[width=\linewidth]{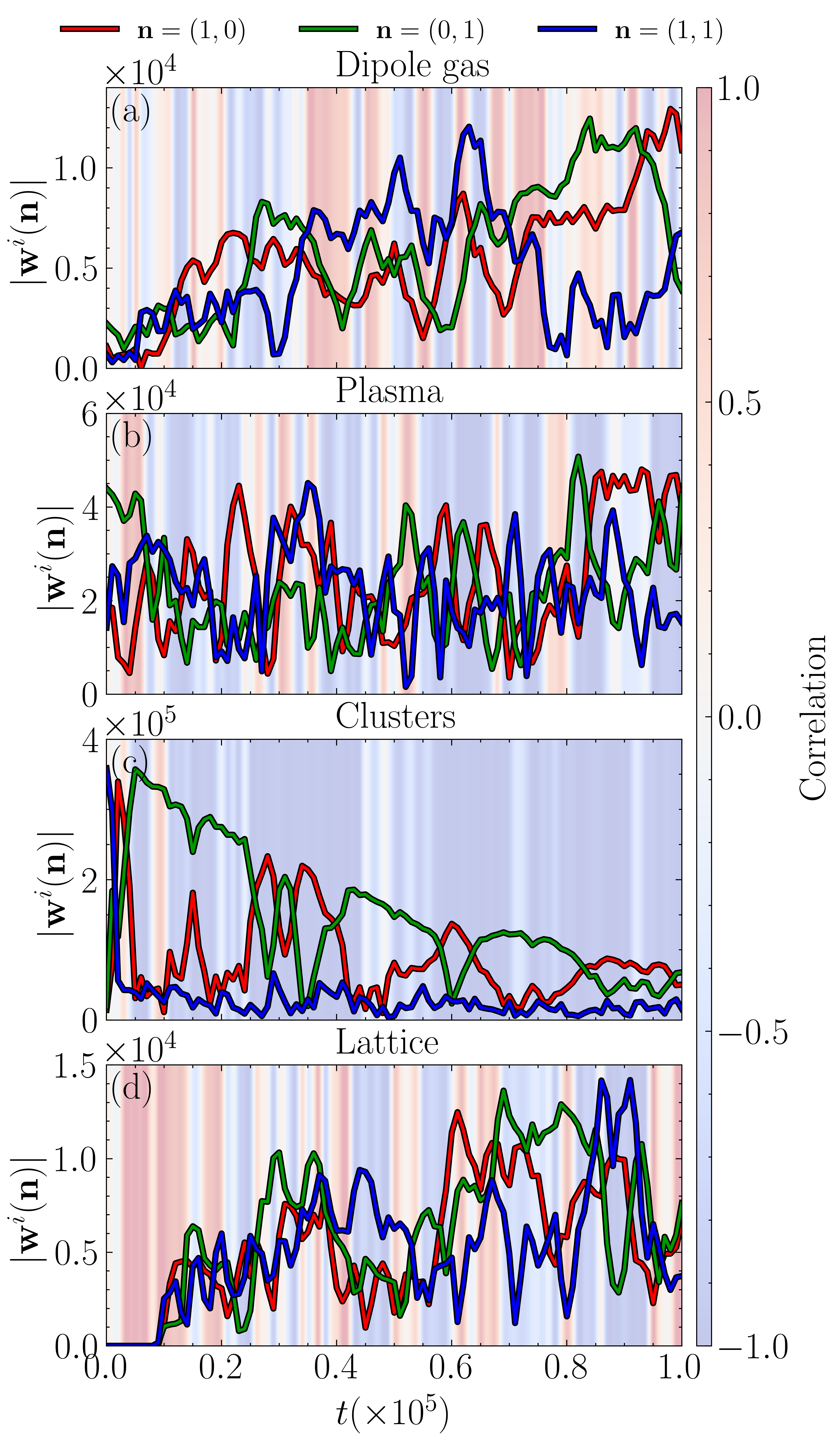}
\caption{The time evolution of three low-lying Fourier modes associated with incompressible component of density weighted velocity field $|w^{i}(n)|$ for the four vortex distributions: (a) dipole, (b) plasma, (c) cluster, and (d) lattice. The correlation between the $n=(0,1)$ and $n=(1,0)$ modes is shown in the background with pseudo-color.  }\label{fig:Wk_final}
\end{figure}
We now turn our attention to a more detailed analysis of the temporal evolution of the amplitudes of three low-lying Fourier modes associated with the incompressible component of the density-weighted velocity field, denoted \( |\mathbf{w}^i(\mathbf{n})| \), as shown in Fig.~\ref{fig:Wk_final}. These modes correspond to specific values of the wavevector index \( \mathbf{n} = (n_x, n_y) \), where the three velocity modes considered here are \((1,0)\), \((0,1)\), and \((1,1)\). The mode with index \((1,0)\) is represented in red, the mode \((0,1)\) in green, and the mode \((1,1)\) in blue, in the respective panels of the figure. The horizontal axis in each panel represents time \( t \), scaled by \( 10^5 \), while the vertical axis shows the magnitude of the mode amplitude, \( |\mathbf{w}^i(\mathbf{n})| \), which quantifies the strength of each mode at each time step.
\begin{figure*}
\centering
\includegraphics[width=\linewidth]{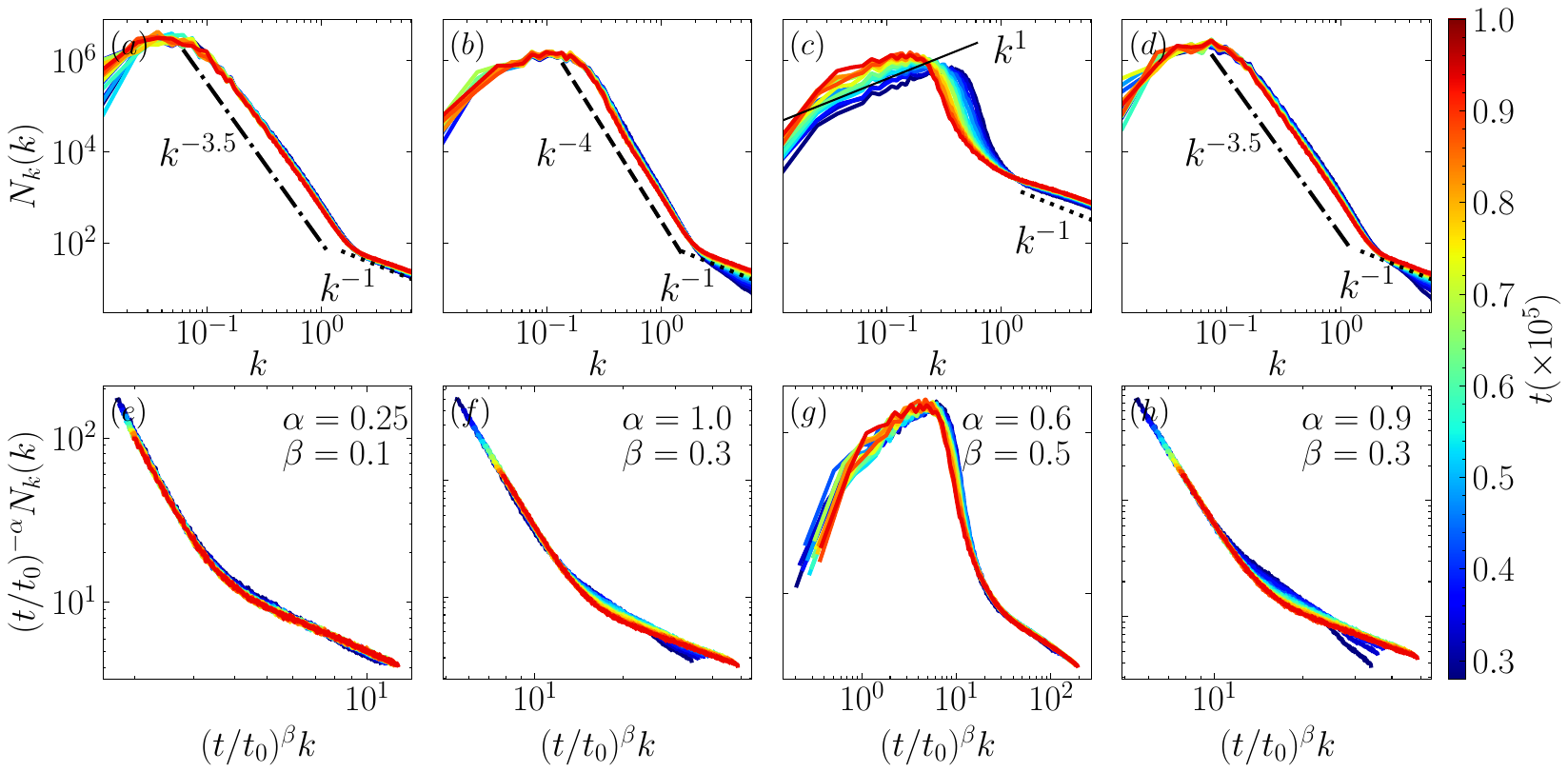}
\caption{ Top row: The $N_k$ spectra as a function of wave number for the different initial conditions: (a) for dipole, (b) for Plasma, (c) for cluster and (d) for lattice.  Bottom row: rescaled $N_k$ spectra [$(t/t_0)^{-\alpha}N_k$] as a function of $(t/t_0)^{\beta}k$  for different initial condition with $t_0 = 100$. For the dipole, plasma and lattice cases, only the high $k$ region ($k>1.0$) is shown in the bottom row,  while for the cluster case, the complete spectra have been rescaled. Pseudo-color is used to represent the spectra at different time. For cluster case the dynamical scaling exponents are $\alpha\simeq 0.6$ and $\beta\simeq0.5$, which ascertain the near equilibrium state. }\label{fig:NK_FSS}
\end{figure*}
In addition to the mode amplitudes, the background of each panel is shaded to represent the instantaneous correlation between the amplitudes of the \(\mathbf{n} = (1,0)\) and \(\mathbf{n} = (0,1)\) modes over time. This correlation is quantified using a rolling Pearson correlation coefficient~\cite{Press:CUP2007} defined as
\begin{equation}
    Corr_{xy}(t_j) = \frac{\sum_{i\in W_j} (x_i -\bar{x})(y_i -\bar{y})}{\sqrt{\sum_{i\in W_j} (x_i -\bar{x})^2}\sqrt{\sum_{i\in W_j}(y_i -\bar{y})^2}}
\end{equation}
where $x(t_j)$ and $y(t_j)$ are two discrete time series, $\bar{x}$ and $\bar{y}$ represent the mean values of $x$ and $y$, respectively, computed over the window $W_j=\{j-w+1,\ldots,j\}$ and $w$ represents the window size. The value of the correlation coefficient, which ranges from \([-1,1]\), is mapped onto a diverging color scale. Red tones correspond to a strong positive correlation, indicating that the amplitudes of the two modes vary together in a synchronized manner (both increase or decrease in unison). Blue tones represent a strong negative correlation, meaning that the two modes tend to evolve in opposite directions (when one increases, the other decreases). White regions, on the other hand, indicate a near-zero correlation, implying no significant linear relationship between the mode amplitudes within the time window under consideration.

The figure presents four stacked panels, each corresponding to a different case under investigation. Our analysis reveals distinct behaviors in the temporal evolution of the mode amplitudes across these different cases. Specifically, we observe no significant long-term correlation or anti-correlation between the modes in any of the cases, except for the cluster case. In the cluster case, a clear and persistent anti-correlation emerges between the \((1,0)\) and \((0,1)\) modes. This anti-correlation suggests that the two modes evolve in opposite directions over time, with the magnitudes of their amplitudes showing an inverse relationship: when the amplitude of one mode increases, the other decreases, and vice versa. This behavior is notably absent in the other cases. Analogous features have also been reported in numerical studies of the Euler equation~\cite{Verma:PRA2022}.

In addition to this anti-correlation, we also note that in the cluster case, the occupancy of the \((1,1)\) mode is significantly lower than that of the \((1,0)\) and \((0,1)\) modes. This means that the \((1,1)\) mode contributes less to the overall dynamics compared to the other two modes in this case, an effect that is not observed in any of the other cases. In contrast, in the dipole and lattice cases, we observe a marked growth in the occupancy of all three modes over time, with the amplitude of each mode increasing steadily. This indicates a more uniform or synchronized evolution of the modes in these cases, in contrast to the behavior observed in the cluster case.

\subsection{Dynamical scalings of particle number spectrum}
In this section, we focus on the dynamical scaling of the particle number spectrum, with particular emphasis on the critical exponents associated with nonthermal fixed points~\cite{Chantesana:PRA2019}. These scaling behaviors are often characterized by power-law distributions, and they provide important insights into the nature of the underlying dynamics of the system as it evolves towards a steady state~\cite {Mikheev:EPJ2023, Madeira:PNAS2024}.

In Fig.~\ref{fig:NK_FSS}, we present the particle number spectra for four distinct initial configurations: dipole, plasma, cluster, and lattice, shown in the top row of the figure. Each color represents the spectrum at a specific time, as indicated by the accompanying colorbar. The analysis reveals a characteristic \( k^{-1} \) scaling in the ultraviolet (UV) regime across all initial configurations, a behavior that is well-known in the literature and corresponds to thermodynamic equipartition of energy modes in the system~\cite{Nazarenko:PDNP2006}. The observation of this universal scaling across all cases strongly suggests that the system tends towards a common thermalized state at small scales. This is consistent with the scaling behavior observed in the UV regime of the compressible kinetic energy spectra, where a similar \( k^{-1} \) scaling is seen. This scaling implies a distribution of energy among modes that is characteristic of thermalized systems, where the energy is evenly distributed among the accessible modes.

However, the infrared (IR) region of the spectra exhibits more complexity and shows marked differences depending on the initial vortex configuration. Previous studies~\cite{Nowak:PRA2012, Karl:NJP2017} have examined the power laws in the particle number spectra for a system of vortices, reporting $k^{-3}$ and $k^{-1}$ scalings, which were attributed to statistically independent vortices and randomly distributed vortex pairs, respectively. We note that our definition of the particle number spectra contains an additional factor of $k$ as compared to ~\cite{Nazarenko:PDNP2006, Nowak:NJP2014, Karl:NJP2017}.

In our study, the spectra for the dipole and lattice configurations exhibit a scaling close to \( k^{-3.5} \) in the IR regime. Since the corresponding vortex configuration lies between those of a randomly distributed vortex gas and a system of randomly distributed, well-separated vortex dipoles, one would expect an exponent between $-1$ and $-3$. The deviation from this expected interval is likely due to contributions from the the compressible component, which remains non-negligible in the IR region. The plasma configuration exhibits an even steeper scaling, approximately \( k^{-4} \), which we attribute to the comparatively smaller number of vortex dipoles present in this case. In stark contrast to these vortex-driven dynamics, the cluster configuration exhibits a different scaling behavior in the IR region, specifically \( k \), which is indicative of particle-number equipartition~\cite{Nazarenko:PDNP2006}. The cluster configuration thus appears to follow a different physical pathway towards equilibrium, one that is driven by an inverse particle cascade which leads to equipartition at small wavenumbers~\cite{Karl:NJP2017}.

The bottom row of Fig.~\ref{fig:NK_FSS} shows the same spectra after they have been rescaled according to a self-similar transformation~\cite{Madeira:PNAS2024}:
\begin{equation}
    N_k(t,k) = \left(\frac{t}{t_0}\right)^\alpha N_k\left(t_0,\left(\frac{t}{t_0}\right)^\beta k\right),
\end{equation}
where \( t_0 \) is a reference time (taken here as \( 10^2 \)). The values of the exponents \( \alpha \) and \( \beta \) for each configuration are indicated in the corresponding subplots. This rescaling procedure is designed to highlight the self-similar evolution of the spectra over time, which is a hallmark of many non-equilibrium systems that approach a self-organized steady state. 

For the dipole, plasma, and lattice cases [Figs.~\ref{fig:NK_FSS}(e), (f), and (h)], the rescaling is applied only to the portion of the spectra corresponding to \( k \gtrsim 1.0 \) in all cases except the cluster case, as this is the region where significant scaling behavior is observed. The extracted scaling exponents for these cases are as follows: \( \alpha \simeq 0.25 \) and \( \beta \simeq 0.1 \) for the dipole case, while both the plasma and lattice cases yield exponents \( \alpha \sim 1.0 \) and \( \beta \sim 0.3 \). These values suggest that, while all three configurations exhibit self-similar scaling, the degree of evolution towards a steady state differs. The dipole case exhibits a less pronounced self-similar behavior compared to the plasma and lattice cases, which have more pronounced scaling behavior with exponents closer to unity. The cluster case, however, displays a distinctly different behavior in its rescaled spectrum, with exponents \( \alpha = 0.6 \) and \( \beta = 0.5 \) across the entire wavenumber range. This suggests that the cluster configuration undergoes a more complex evolution towards equilibration, where both large- and small-scale modes evolve simultaneously in a self-similar fashion. Our exponents agree well with those reported in previous studies~\cite{Madeira:PNAS2024}.

This distinct behavior in the cluster configuration highlights the role of initial conditions in determining the long-term dynamics of the system. The simultaneous participation of both large- and small-scale modes in the self-similar evolution indicates that the cluster configuration follows a more intricate trajectory towards a steady state, one that may be influenced by different physical processes than those governing the dipole, plasma, and lattice cases. Thus, while all configurations exhibit self-similar evolution in the high-$k$ regime, the cluster case reveals the possibility of more complex, multiscale dynamics, which may be a key feature of certain non-equilibrium systems.

\section{Summary and Conclusion}
\label{sec:con} 
In this paper, we have numerically investigated the evolution of four distinct vortex configurations—dipole, plasma, cluster, and lattice—using the two-dimensional mean-field Gross-Pitaevskii equation. By analyzing the decay of vortex number over time, we were able to draw important insights into the dynamics of these configurations as they evolve towards equilibrium. Our results demonstrate that the cluster configuration exhibits the fastest relaxation time, reaching equilibrium significantly earlier than the dipole, plasma, and lattice configurations. This rapid equilibration is accompanied by distinctive scaling behaviors in both the incompressible and compressible spectra.

For the incompressible spectrum, the dipole (over a narrow band of wavenumbers) and cluster configurations initially exhibit Kolmogorov-like scaling ($\varepsilon^i_{kin}\sim k^{-5/3}$), which is characteristic of turbulence in classical fluids. At later times, both configurations transition to a Vinen-like scaling characterized by ($\varepsilon^i_{kin}\sim k^{-1}$), however, the Kolmogorov spectra re-emerges in the cluster case at still later times. The remaining configurations predominantly show Vinen-like scaling throughout their evolution.
In the compressible spectrum, the cluster configuration shows a scaling behavior of $k$ signaling that all modes have fully equilibrated, whereas for the other configurations, modes thermalize only above a critical wave number. This difference in mode equilibration is crucial for understanding the dynamics of quantum turbulence, as it reveals that the cluster configuration achieves a more homogeneous distribution of energy across different scales. 

Further, our transfer function analysis provides insight into the nature of the equilibrium state for each configuration. The cluster case shows a perfectly Gaussian distribution for all modes, reflecting a state of complete thermalization, while the other configurations exhibit skewed Gaussian or exponential distributions, suggesting a less uniform energy distribution and slower relaxation toward equilibrium. These findings highlight the cluster configuration as the most efficient in reaching a thermodynamically stable state, making it particularly relevant for studies of quantum turbulence in Bose-Einstein condensates and superfluids.

Finally, our particle number spectra analysis reinforces the notion that the cluster case reaches dynamic scaling closer to equilibrium compared to the other initial configurations which is further reinforced through the dynamical exponent of the particle number spectra. This emphasizes the potential of the cluster configuration for studies aimed at controlling and manipulating quantum turbulence, as its faster relaxation could be advantageous in experiments that require rapid stabilization of vortex dynamics.

Overall, our numerical investigation provides valuable insights into the different pathways through which vortex configurations evolve toward equilibrium in quantum fluids. We have analyzed the relaxation of various initial vortex configurations into both equilibrium and non-equilibrium states, specifically focusing on the decay of the turbulent state in the condensate. Extending this work to the steady-state turbulent regime produced by the simultaneous presence of forcing and damping in the GPE, as explored in Refs.~\cite{Nazarenko:PDNP2006, Proment:PRA2009, Proment:PDNP2012, Zhu:PRL2023}, would be an interesting avenue for future research. Another natural extension is to study the role of dimensionality~\cite{Yang:arXiv2025} and vortex density~\cite{Kanai:PRA2025} on the evolution and decay of different vortex configurations. It would also be instructive to conduct similar analyses in multi-component BECs~\cite{Wheeler:EPL2021, Madeira:AVSQS2020} and examine the effect of intercomponent interactions and additional degrees of freedom. Incorporating beyond-mean-field corrections would likewise offer important insights into regimes where quantum fluctuations play an important role~\cite{Jha:PRF2025} and may help bridge the gap between simulations and experimental observations. A future compelling direction of research area to explore similar ideas would be to examine analogous phenomena in dipolar BECs, where long-range interactions introduce additional complexity and can generate a nontrivial dependence on initial conditions~\cite{Rasch:PRA2025, Lahaye:ROPIP2009, Chomaz:ROPIP2022}.

\section{Acknowledgments}
\label{sec:ack}
We would like to thank Mahendra K. Verma and Sachin Singh Rawat for discussions and fruitful suggestions at the initial stage of our work. M.T. acknowledges the
support from JSPS KAKENHI Grants No. JP22H05131, JP22H05139, and JP23K03305. S.K.J acknowledges financial support from the University Grants Commission - Council of Scientific and Industrial Research (UGC-CSIR), India.
\bibliography{references.bib}

@article{Nore:PRL1997,
  title = {Kolmogorov Turbulence in Low-Temperature Superflows},
  author = {Nore, C. and Abid, M. and Brachet, M. E.},
  journal = {Phys. Rev. Lett.},
  volume = {78},
  issue = {20},
  pages = {3896--3899},
  numpages = {0},
  year = {1997},
  month = {May},
  publisher = {American Physical Society},
  doi = {10.1103/PhysRevLett.78.3896},
  url = {https://link.aps.org/doi/10.1103/PhysRevLett.78.3896}
}

@article{Nore:POF1997,
    author = {Nore, C. and Abid, M. and Brachet, M. E.},
    title = {Decaying Kolmogorov turbulence in a model of superflow},
    journal = {Phys. Fluids},
    volume = {9},
    number = {9},
    pages = {2644-2669},
    year = {1997},
    month = {09},
    issn = {1070-6631},
    doi = {10.1063/1.869473},
    url = {https://doi.org/10.1063/1.869473},
}

@article{Berloff:PRA2002,
  title = {Scenario of strongly nonequilibrated {{{{Bose-Einstein}}}} condensation},
  author = {Berloff, Natalia G. and Svistunov, Boris V.},
  journal = {Phys. Rev. A},
  volume = {66},
  issue = {1},
  pages = {013603},
  numpages = {7},
  year = {2002},
  month = {Jul},
  publisher = {American Physical Society},
  doi = {10.1103/PhysRevA.66.013603},
  url = {https://link.aps.org/doi/10.1103/PhysRevA.66.013603}
}

@article{Kobayashi:PRL2005,
  title = {Kolmogorov Spectrum of Superfluid Turbulence: Numerical Analysis of the Gross-Pitaevskii Equation with a Small-Scale Dissipation},
  author = {Kobayashi, Michikazu and Tsubota, Makoto},
  journal = {Phys. Rev. Lett.},
  volume = {94},
  issue = {6},
  pages = {065302},
  numpages = {4},
  year = {2005},
  month = {Feb},
  publisher = {American Physical Society},
  doi = {10.1103/PhysRevLett.94.065302},
  url = {https://link.aps.org/doi/10.1103/PhysRevLett.94.065302}
}

@article{Nazarenko:PDNP2006,
    title = {Wave turbulence and vortices in Bose–Einstein condensation},
    journal = {Physica D: Nonlin. Phen.},
    volume = {219},
    number = {1},
    pages = {1-12},
    year = {2006},
    issn = {0167-2789},
    doi = {https://doi.org/10.1016/j.physd.2006.05.007},
    url = {https://www.sciencedirect.com/science/article/pii/S0167278906001758},
    author = {Sergey Nazarenko and Miguel Onorato},
}

@Article{Nazarenko:JLTP2006,
  author    = {Sergey Nazarenko and Miguel Onorato},
  journal   = {J. Low Temp. Phys.},
  title     = {Freely decaying Turbulence and {{{{{Bose-Einstein}}}}} Condensation in {Gross-Pitaevski} Model},
  year      = {2006},
  month     = {dec},
  number    = {1-2},
  pages     = {31--46},
  volume    = {146},
  doi       = {10.1007/s10909-006-9271-z},
  publisher = {Springer Science and Business Media {LLC}},
}

@article{Nazarenko:JLTP2007,
    title={Freely decaying Turbulence and Bose–Einstein Condensation in Gross–Pitaevski Model},
    author={Nazarenko, Sergey and Onorato, Miguel},
    journal={J. Low Temp. Phys.},
    volume={146},
    number={1},
    pages={31--46},
    year={2007},
    publisher={Springer},
    doi={10.1007/s10909-006-9271-z}
}

@article{Proment:PRA2009,
  title = {Quantum turbulence cascades in the Gross-Pitaevskii model},
  author = {Proment, Davide and Nazarenko, Sergey and Onorato, Miguel},
  journal = {Phys. Rev. A},
  volume = {80},
  issue = {5},
  pages = {051603},
  numpages = {4},
  year = {2009},
  month = {Nov},
  publisher = {American Physical Society},
  doi = {10.1103/PhysRevA.80.051603},
  url = {https://link.aps.org/doi/10.1103/PhysRevA.80.051603}
}

@article{Berges:NPB2009,
title = {Nonthermal fixed points and the functional renormalization group},
author = {Jürgen Berges and Gabriele Hoffmeister},
journal = {Nucl. Phys. B},
volume = {813},
number = {3},
pages = {383-407},
year = {2009},
issn = {0550-3213},
doi = {10.1016/j.nuclphysb.2008.12.017},
url = {https://www.sciencedirect.com/science/article/pii/S0550321308007219}
}

@article{Henn:PRL2009,
  title = {Emergence of Turbulence in an Oscillating {{{{Bose-Einstein}}}} Condensate},
  author = {Henn, E. A. L. and Seman, J. A. and Roati, G. and Magalh\~aes, K. M. F. and Bagnato, V. S.},
  journal = {Phys. Rev. Lett.},
  volume = {103},
  issue = {4},
  pages = {045301},
  numpages = {4},
  year = {2009},
  month = {Jul},
  publisher = {American Physical Society},
  doi = {10.1103/PhysRevLett.103.045301},
  url = {https://link.aps.org/doi/10.1103/PhysRevLett.103.045301}
}

@article{Lahaye:ROPIP2009,
    doi = {10.1088/0034-4885/72/12/126401},
    url = {https://doi.org/10.1088/0034-4885/72/12/126401},
    year = {2009},
    month = {nov},
    publisher = {},
    volume = {72},
    number = {12},
    pages = {126401},
    author = {Lahaye, T and Menotti, C and Santos, L and Lewenstein, M and Pfau, T},
    title = {The physics of dipolar bosonic quantum gases},
    journal = {Rep. Prog. Phys.},
}

@article{Numasato:JLTP2010,
  title={Possibility of inverse energy cascade in two-dimensional quantum turbulence},
  author={Numasato, Ryu and Tsubota, Makoto},
  journal={J. Low Temp. Phys.},
  volume={158},
  number={3},
  pages={415--421},
  year={2010},
  publisher={Springer}
}

@article{Numasato:PRA2010,
  title = {Direct energy cascade in two-dimensional compressible quantum turbulence},
  author = {Numasato, Ryu and Tsubota, Makoto and L'vov, Victor S.},
  journal = {Phys. Rev. A},
  volume = {81},
  issue = {6},
  pages = {063630},
  numpages = {12},
  year = {2010},
  month = {Jun},
  publisher = {American Physical Society},
  doi = {10.1103/PhysRevA.81.063630},
  url = {https://link.aps.org/doi/10.1103/PhysRevA.81.063630}
}

@article{White:PRL2010,
  title = {Nonclassical Velocity Statistics in a Turbulent Atomic {{{{Bose-Einstein}}}} Condensate},
  author = {White, A. C. and Barenghi, C. F. and Proukakis, N. P. and Youd, A. J. and Wacks, D. H.},
  journal = {Phys. Rev. Lett.},
  volume = {104},
  issue = {7},
  pages = {075301},
  numpages = {4},
  year = {2010},
  month = {Feb},
  publisher = {American Physical Society},
  doi = {10.1103/PhysRevLett.104.075301},
  url = {https://link.aps.org/doi/10.1103/PhysRevLett.104.075301}
}

@article{Nowak:PRB2011,
  title = {Superfluid turbulence: Nonthermal fixed point in an ultracold Bose gas},
  author = {Nowak, Boris and Sexty, D\'enes and Gasenzer, Thomas},
  journal = {Phys. Rev. B},
  volume = {84},
  issue = {2},
  pages = {020506},
  numpages = {4},
  year = {2011},
  month = {Jul},
  publisher = {American Physical Society},
  doi = {10.1103/PhysRevB.84.020506},
  url = {https://link.aps.org/doi/10.1103/PhysRevB.84.020506}
}

@article{Bradley:PRX2012,
  title = {Energy Spectra of Vortex Distributions in Two-Dimensional Quantum Turbulence},
  author = {Bradley, Ashton S. and Anderson, Brian P.},
  journal = {Phys. Rev. X},
  volume = {2},
  issue = {4},
  pages = {041001},
  numpages = {20},
  year = {2012},
  month = {Oct},
  publisher = {American Physical Society},
  doi = {10.1103/PhysRevX.2.041001},
  url = {https://link.aps.org/doi/10.1103/PhysRevX.2.041001}
}

@article{Berges:PRL2012,
    title = {{{{{Bose-Einstein}}}} Condensation in Relativistic Field Theories Far from Equilibrium},
    author = {Berges, J\"urgen and Sexty, D\'enes},
    journal = {Phys. Rev. Lett.},
    volume = {108},
    issue = {16},
    pages = {161601},
    numpages = {4},
    year = {2012},
    month = {Apr},
    publisher = {American Physical Society},
    doi = {10.1103/PhysRevLett.108.161601},
    url = {https://link.aps.org/doi/10.1103/PhysRevLett.108.161601}
}

@article{Nowak:PRA2012,
    title = {Nonthermal fixed points, vortex statistics, and superfluid turbulence in an ultracold Bose gas},
    author = {Nowak, Boris and Schole, Jan and Sexty, D\'enes and Gasenzer, Thomas},
    journal = {Phys. Rev. A},
    volume = {85},
    issue = {4},
    pages = {043627},
    numpages = {19},
    year = {2012},
    month = {Apr},
    publisher = {American Physical Society},
    doi = {10.1103/PhysRevA.85.043627},
    url = {https://link.aps.org/doi/10.1103/PhysRevA.85.043627}
}

@article{Proment:PDNP2012,
    title = {Sustained turbulence in the three-dimensional Gross–Pitaevskii model},
    journal = {Physica D: Nonlin. Phen.},
    volume = {241},
    number = {3},
    pages = {304-314},
    year = {2012},
    note = {Special Issue on Small Scale Turbulence},
    issn = {0167-2789},
    doi = {https://doi.org/10.1016/j.physd.2011.06.007},
    url = {https://www.sciencedirect.com/science/article/pii/S0167278911001564},
    author = {Davide Proment and Sergey Nazarenko and Miguel Onorato},
}

@article{Reeves:PRA2012,
  title = {Classical and quantum regimes of two-dimensional turbulence in trapped {{{{Bose-Einstein}}}} condensates},
  author = {Reeves, M. T. and Anderson, B. P. and Bradley, A. S.},
  journal = {Phys. Rev. A},
  volume = {86},
  issue = {5},
  pages = {053621},
  numpages = {11},
  year = {2012},
  month = {Nov},
  publisher = {American Physical Society},
  doi = {10.1103/PhysRevA.86.053621},
  url = {https://link.aps.org/doi/10.1103/PhysRevA.86.053621}
}

@article{White:PRA2012,
  title = {Creation and characterization of vortex clusters in atomic {{{{Bose-Einstein}}}} condensates},
  author = {White, Angela C. and Barenghi, Carlo F. and Proukakis, Nick P.},
  journal = {Phys. Rev. A},
  volume = {86},
  issue = {1},
  pages = {013635},
  numpages = {8},
  year = {2012},
  month = {Jul},
  publisher = {American Physical Society},
  doi = {10.1103/PhysRevA.86.013635},
  url = {https://link.aps.org/doi/10.1103/PhysRevA.86.013635}
}

@article{Neely:PRL2013,
  title = {Characteristics of Two-Dimensional Quantum Turbulence in a Compressible Superfluid},
  author = {Neely, T. W. and Bradley, A. S. and Samson, E. C. and Rooney, S. J. and Wright, E. M. and Law, K. J. H. and Carretero-Gonz\'alez, R. and Kevrekidis, P. G. and Davis, M. J. and Anderson, B. P.},
  journal = {Phys. Rev. Lett.},
  volume = {111},
  issue = {23},
  pages = {235301},
  numpages = {6},
  year = {2013},
  month = {Dec},
  publisher = {American Physical Society},
  doi = {10.1103/PhysRevLett.111.235301},
  url = {https://link.aps.org/doi/10.1103/PhysRevLett.111.235301}
}

@article{Reeves:PRL2013,
  title = {Inverse Energy Cascade in Forced Two-Dimensional Quantum Turbulence},
  author = {Reeves, Matthew T. and Billam, Thomas P. and Anderson, Brian P. and Bradley, Ashton S.},
  journal = {Phys. Rev. Lett.},
  volume = {110},
  issue = {10},
  pages = {104501},
  numpages = {5},
  year = {2013},
  month = {Mar},
  publisher = {American Physical Society},
  doi = {10.1103/PhysRevLett.110.104501},
  url = {https://link.aps.org/doi/10.1103/PhysRevLett.110.104501}
}

@article{Barenghi:PNAS2014,
author = {Carlo F. Barenghi  and Ladislav Skrbek  and Katepalli R. Sreenivasan },
title = {Introduction to quantum turbulence},
journal = {Proc. Nat. Acad. Sci.},
volume = {111},
number = {supplement\_1},
pages = {4647-4652},
year = {2014},
doi = {10.1073/pnas.1400033111},
URL = {https://www.pnas.org/doi/abs/10.1073/pnas.1400033111},
}

@article{Billam:PRL2014,
  title = {Onsager-Kraichnan Condensation in Decaying Two-Dimensional Quantum Turbulence},
  author = {Billam, T. P. and Reeves, M. T. and Anderson, B. P. and Bradley, A. S.},
  journal = {Phys. Rev. Lett.},
  volume = {112},
  issue = {14},
  pages = {145301},
  numpages = {6},
  year = {2014},
  month = {Apr},
  publisher = {American Physical Society},
  doi = {10.1103/PhysRevLett.112.145301},
  url = {https://link.aps.org/doi/10.1103/PhysRevLett.112.145301}
}

@article{Kwon:PRA2014,
  title = {Relaxation of superfluid turbulence in highly oblate {{{{Bose-Einstein}}}} condensates},
  author = {Kwon, Woo Jin and Moon, Geol and Choi, Jae-yoon and Seo, Sang Won and Shin, Yong-il},
  journal = {Phys. Rev. A},
  volume = {90},
  issue = {6},
  pages = {063627},
  numpages = {6},
  year = {2014},
  month = {Dec},
  publisher = {American Physical Society},
  doi = {10.1103/PhysRevA.90.063627},
  url = {https://link.aps.org/doi/10.1103/PhysRevA.90.063627}
}

@article{Simula:PRL2014,
  title={Emergence of order from turbulence in an isolated planar superfluid},
  author={Simula, Tapio and Davis, Matthew J and Helmerson, Kristian},
  journal={Phys. Rev. Lett.},
  volume={113},
  number={16},
  pages={165302},
  year={2014},
  publisher={APS}
}

@article{Nowak:NJP2014,
doi = {10.1088/1367-2630/16/9/093052},
url = {https://doi.org/10.1088/1367-2630/16/9/093052},
year = {2014},
month = {sep},
publisher = {IOP Publishing},
volume = {16},
number = {9},
pages = {093052},
author = {Nowak, Boris and Schole, Jan and Gasenzer, Thomas},
title = {Universal dynamics on the way to thermalization},
journal = {New J. Phys.},
}

@article{Billam:PRA2015,
  title = {Spectral energy transport in two-dimensional quantum vortex dynamics},
  author = {Billam, T. P. and Reeves, M. T. and Bradley, A. S.},
  journal = {Phys. Rev. A},
  volume = {91},
  issue = {2},
  pages = {023615},
  numpages = {6},
  year = {2015},
  month = {Feb},
  publisher = {American Physical Society},
  doi = {10.1103/PhysRevA.91.023615},
  url = {https://link.aps.org/doi/10.1103/PhysRevA.91.023615}
}

@article{Stagg:PRA2015,
  title = {Generation and decay of two-dimensional quantum turbulence in a trapped {{{{Bose-Einstein}}}} condensate},
  author = {Stagg, G. W. and Allen, A. J. and Parker, N. G. and Barenghi, C. F.},
  journal = {Phys. Rev. A},
  volume = {91},
  issue = {1},
  pages = {013612},
  numpages = {6},
  year = {2015},
  month = {Jan},
  publisher = {American Physical Society},
  doi = {10.1103/PhysRevA.91.013612},
  url = {https://link.aps.org/doi/10.1103/PhysRevA.91.013612}
}

@article{Orioli:PRD2015,
  title = {Universal self-similar dynamics of relativistic and nonrelativistic field theories near nonthermal fixed points},
  author = {Pi\~neiro Orioli, Asier and Boguslavski, Kirill and Berges, J\"urgen},
  journal = {Phys. Rev. D},
  volume = {92},
  issue = {2},
  pages = {025041},
  numpages = {25},
  year = {2015},
  month = {Jul},
  publisher = {American Physical Society},
  doi = {10.1103/PhysRevD.92.025041},
  url = {https://link.aps.org/doi/10.1103/PhysRevD.92.025041}
}

@article{Groszek:PRA2016,
  title={Onsager vortex formation in {{{{Bose-Einstein}}}} condensates in two-dimensional power-law traps},
  author={Groszek, Andrew J and Simula, Tapio P and Paganin, David M and Helmerson, Kristian},
  journal={Phys. Rev. A},
  volume={93},
  number={4},
  pages={043614},
  year={2016},
  publisher={APS}
}

@article{Cidrim:PRA2016,
  title = {Controlled polarization of two-dimensional quantum turbulence in atomic {Bose-Einstein} condensates},
  author = {Cidrim, A. and dos Santos, F. E. A. and Galantucci, L. and Bagnato, V. S. and Barenghi, C. F.},
  journal = {Phys. Rev. A},
  volume = {93},
  issue = {3},
  pages = {033651},
  numpages = {8},
  year = {2016},
  month = {Mar},
  publisher = {American Physical Society},
  doi = {10.1103/PhysRevA.93.033651},
  url = {https://link.aps.org/doi/10.1103/PhysRevA.93.033651}
}

@article{Karl:NJP2017,
    doi = {10.1088/1367-2630/aa7eeb},
    url = {https://doi.org/10.1088/1367-2630/aa7eeb},
    year = {2017},
    month = {sep},
    publisher = {IOP Publishing},
    volume = {19},
    number = {9},
    pages = {093014},
    author = {Karl, Markus and Gasenzer, Thomas},
    title = {Strongly anomalous non-thermal fixed point in a quenched two-dimensional Bose gas},
    journal = {New J. Phys.},
}

@article{Baggaley:PRA2018,
  title = {Decay of homogeneous two-dimensional quantum turbulence},
  author = {Baggaley, Andrew W. and Barenghi, Carlo F.},
  journal = {Phys. Rev. A},
  volume = {97},
  issue = {3},
  pages = {033601},
  numpages = {5},
  year = {2018},
  month = {Mar},
  publisher = {American Physical Society},
  doi = {10.1103/PhysRevA.97.033601},
  url = {https://link.aps.org/doi/10.1103/PhysRevA.97.033601}
}

@article{Groszek:PRA2018,
  title = {Motion of vortices in inhomogeneous {{{{Bose-Einstein}}}} condensates},
  author = {Groszek, Andrew J. and Paganin, David M. and Helmerson, Kristian and Simula, Tapio P.},
  journal = {Phys. Rev. A},
  volume = {97},
  issue = {2},
  pages = {023617},
  numpages = {12},
  year = {2018},
  month = {Feb},
  publisher = {American Physical Society},
  doi = {10.1103/PhysRevA.97.023617},
  url = {https://link.aps.org/doi/10.1103/PhysRevA.97.023617}
}

@article{Valani:NJOP2018,
doi = {10.1088/1367-2630/aac0bb},
url = {https://doi.org/10.1088/1367-2630/aac0bb},
year = {2018},
month = {may},
publisher = {IOP Publishing},
volume = {20},
number = {5},
pages = {053038},
author = {Valani, Rahil N and Groszek, Andrew J and Simula, Tapio P},
title = {Einstein–Bose condensation of Onsager vortices},
journal = {New J. Phys.}
}

@article{Chantesana:PRA2019,
  title = {Kinetic theory of nonthermal fixed points in a Bose gas},
  author = {Chantesana, Isara and Orioli, Asier Pi\~neiro and Gasenzer, Thomas},
  journal = {Phys. Rev. A},
  volume = {99},
  issue = {4},
  pages = {043620},
  numpages = {46},
  year = {2019},
  month = {Apr},
  publisher = {American Physical Society},
  doi = {10.1103/PhysRevA.99.043620},
  url = {https://link.aps.org/doi/10.1103/PhysRevA.99.043620}
}

@article{Gauthier:Sci2019,
  author = {Guillaume Gauthier  and Matthew T. Reeves  and Xiaoquan Yu  and Ashton S. Bradley  and Mark A. Baker  and Thomas A. Bell  and Halina Rubinsztein-Dunlop  and Matthew J. Davis  and Tyler W. Neely },
  title = {Giant vortex clusters in a two-dimensional quantum fluid},
  journal = {Science},
  volume = {364},
  number = {6447},
  pages = {1264-1267},
  year = {2019},
  doi = {10.1126/science.aat5718}
}

@article{Johnstone:Sci2019,
  author = {Shaun P. Johnstone  and Andrew J. Groszek  and Philip T. Starkey  and Christopher J. Billington  and Tapio P. Simula  and Kristian Helmerson },
  title = {Evolution of large-scale flow from turbulence in a two-dimensional superfluid},
  journal = {Science},
  volume = {364},
  number = {6447},
  pages = {1267-1271},
  year = {2019},
  doi = {10.1126/science.aat5793},
}

@article{Navon:Sci2019,
  author = {Nir Navon  and Christoph Eigen  and Jinyi Zhang  and Raphael Lopes  and Alexander L. Gaunt  and Kazuya Fujimoto  and Makoto Tsubota  and Robert P. Smith  and Zoran Hadzibabic },
  title = {Synthetic dissipation and cascade fluxes in a turbulent quantum gas},
  journal = {Science},
  volume = {366},
  number = {6463},
  pages = {382-385},
  year = {2019},
  doi = {10.1126/science.aau6103},
  URL = {https://www.science.org/doi/abs/10.1126/science.aau6103},
}

@article{Groszek:SPP2020,
    title={{Decaying quantum turbulence in a two-dimensional {{{{Bose-Einstein}}}} condensate at finite temperature}},
	author={Andrew J. Groszek and Matthew J. Davis and Tapio P. Simula},
	journal={SciPost Phys.},
	volume={8},
	pages={039},
	year={2020},
	publisher={SciPost},
	doi={10.21468/SciPostPhys.8.3.039},
	url={https://scipost.org/10.21468/SciPostPhys.8.3.039},
}

@article{Madeira:AVSQS2020,
  title={Quantum turbulence in Bose--Einstein condensates: Present status and new challenges ahead},
  author={Madeira, Lucas and Cidrim, Andr{\'e} and Hemmerling, Michal and Caracanhas, M{\^o}nica Andrioli and dos Santos, FEA and Bagnato, Vanderlei Salvador},
  journal={AVS Quantum Sci.},
  volume={2},
  number={3},
  pages={035901},
  year={2020},  
  doi = {10.1116/5.0016751},
  publisher={American Vacuum Society}
}

@article{Wheeler:EPL2021,
    doi = {10.1209/0295-5075/ac2c53},
    url = {https://doi.org/10.1209/0295-5075/ac2c53},
    year = {2021},
    month = {oct},
    publisher = {EDP Sciences, IOP Publishing and Società Italiana di Fisica},
    volume = {135},
    number = {3},
    pages = {30004},
    author = {Wheeler, M. T. and Salman, H. and Borgh, M. O.},
    title = {Relaxation dynamics of half-quantum vortices in a two-dimensional two-component {{{{Bose-Einstein}}}} condensate},
    journal = {Europhys. Lett.},
}

@article{Bradley:PRA2022,
  title = {Spectral analysis for compressible quantum fluids},
  author = {Bradley, Ashton S. and Kumar, R. Kishor and Pal, Sukla and Yu, Xiaoquan},
  journal = {Phys. Rev. A},
  volume = {106},
  issue = {4},
  pages = {043322},
  numpages = {15},
  year = {2022},
  month = {Oct},
  publisher = {American Physical Society},
  doi = {10.1103/PhysRevA.106.043322},
  url = {https://link.aps.org/doi/10.1103/PhysRevA.106.043322}
}

@article{Chomaz:ROPIP2022,
    doi = {10.1088/1361-6633/aca814},
    url = {https://doi.org/10.1088/1361-6633/aca814},
    year = {2022},
    month = {dec},
    publisher = {IOP Publishing},
    volume = {86},
    number = {2},
    pages = {026401},
    author = {Chomaz, Lauriane and Ferrier-Barbut, Igor and Ferlaino, Francesca and Laburthe-Tolra, Bruno and Lev, Benjamin L and Pfau, Tilman},
    title = {Dipolar physics: a review of experiments with magnetic quantum gases},
    journal = {Rep. Prog. Phys.},
}

@article{Reeves:PRX2022,
  title = {Turbulent Relaxation to Equilibrium in a Two-Dimensional Quantum Vortex Gas},
  author = {Reeves, Matthew T. and Goddard-Lee, Kwan and Gauthier, Guillaume and Stockdale, Oliver R. and Salman, Hayder and Edmonds, Timothy and Yu, Xiaoquan and Bradley, Ashton S. and Baker, Mark and Rubinsztein-Dunlop, Halina and Davis, Matthew J. and Neely, Tyler W.},
  journal = {Phys. Rev. X},
  volume = {12},
  issue = {1},
  pages = {011031},
  numpages = {18},
  year = {2022},
  month = {Feb},
  publisher = {American Physical Society},
  doi = {10.1103/PhysRevX.12.011031},
  url = {https://link.aps.org/doi/10.1103/PhysRevX.12.011031}
}

@article{Verma:PRA2022,
  title = {Hydrodynamic entropy and emergence of order in two-dimensional Euler turbulence},
  author = {Verma, Mahendra K. and Chatterjee, Soumyadeep},
  journal = {Phys. Rev. Fluids},
  volume = {7},
  issue = {11},
  pages = {114608},
  numpages = {13},
  year = {2022},
  month = {Nov},
  publisher = {American Physical Society},
  doi = {10.1103/PhysRevFluids.7.114608},
  url = {https://link.aps.org/doi/10.1103/PhysRevFluids.7.114608}
}

@article{Mikheev:EPJ2023,
  title={Universal dynamics and non-thermal fixed points in quantum fluids far from equilibrium},
  author={Mikheev, Aleksandr N. and Siovitz, Ido and Gasenzer, Thomas},
  journal={Eur. Phys. J. Spec. Top.},
  volume={232},
  number={20},
  pages={3393--3415},
  year={2023},
  doi={10.1140/epjs/s11734-023-00974-7}
}

@article{Zhu:PRL2023,
  title = {Direct and Inverse Cascades in Turbulent {{{{Bose-Einstein}}}} Condensates},
  author = {Zhu, Ying and Semisalov, Boris and Krstulovic, Giorgio and Nazarenko, Sergey},
  journal = {Phys. Rev. Lett.},
  volume = {130},
  issue = {13},
  pages = {133001},
  numpages = {6},
  year = {2023},
  month = {Mar},
  publisher = {American Physical Society},
  doi = {10.1103/PhysRevLett.130.133001},
  url = {https://link.aps.org/doi/10.1103/PhysRevLett.130.133001}
}

@article{Madeira:PNAS2024,
author = {Lucas Madeira  and Arnol D. García-Orozco  and Michelle A. Moreno-Armijos  and Amilson R. Fritsch  and Vanderlei S. Bagnato },
title = {Universal scaling in far-from-equilibrium quantum systems: An equivalent differential approach},
journal = {Proc. Nat. Acad. Sci.},
volume = {121},
number = {30},
pages = {e2404828121},
year = {2024},
doi = {10.1073/pnas.2404828121},
URL = {https://www.pnas.org/doi/abs/10.1073/pnas.2404828121},
}

@article{Jha:PRF2025,
  title = {Energy spectra and fluxes of two-dimensional turbulent quantum droplets},
  author = {Jha, Shawan Kumar and Verma, Mahendra K. and Mistakidis, S. I. and Mishra, Pankaj Kumar},
  journal = {Phys. Rev. Fluids},
  volume = {10},
  issue = {6},
  pages = {064618},
  numpages = {26},
  year = {2025},
  month = {Jun},
  publisher = {American Physical Society},
  doi = {10.1103/pj6n-3w6p},
  url = {https://link.aps.org/doi/10.1103/pj6n-3w6p}
}

@article{Rawat:CPC2025,
  title={{quTARANG: A High-performance computing Python package to study turbulence using the Gross-Pitaevskii equation}},
  author={Rawat, Sachin Singh and Jha, Shawan Kumar and Verma, Mahendra Kumar and Mishra, Pankaj Kumar},
  journal={Comput. Phys. Commun.},
  volume={315},
  pages={109725},
  doi={10.1016/j.cpc.2025.109725},
  year={2025}
}

@article{Sreenivasan:ARCMP2025,
   author = "Sreenivasan, Katepalli R. and Schumacher, Jörg",
   title = "What Is the Turbulence Problem, and When May We Regard It as Solved?", 
   journal= "Ann. Rev. Cond. Matt. Phys.",
   year = "2025",
   volume = "16",
   number = "Volume 16, 2025",
   pages = "121-143",
   doi = "10.1146/annurev-conmatphys-031620-095842",
   publisher = "Annual Reviews",
   issn = "1947-5462",
   type = "Journal Article",
  }

@article{Kanai:PRA2025,
  title = {Dynamical crossover of vortex-pair annihilation in two-dimensional quantum turbulence},
  author = {Kanai, Toshiaki and Zhang, Chuanwei},
  journal = {Phys. Rev. A},
  volume = {112},
  issue = {3},
  pages = {033305},
  numpages = {8},
  year = {2025},
  month = {Sep},
  publisher = {American Physical Society},
  doi = {10.1103/6zfl-dc7r},
  url = {https://link.aps.org/doi/10.1103/6zfl-dc7r}
}

@article{Rasch:PRA2025,
  title = {Anomalous nonthermal fixed point in a quasi-two-dimensional dipolar Bose gas},
  author = {Rasch, Niklas and Chomaz, Lauriane and Gasenzer, Thomas},
  journal = {Phys. Rev. A},
  volume = {112},
  issue = {5},
  pages = {053310},
  numpages = {20},
  year = {2025},
  month = {Nov},
  publisher = {American Physical Society},
  doi = {10.1103/x2rj-ptgy},
  url = {https://link.aps.org/doi/10.1103/x2rj-ptgy}
}

@article{Yang:arXiv2025,
  title={Quantum Turbulence Across Dimensions: Crossover from two-to three-dimension},
  author={Yang, Weican and Wang, Xin and Tsubota, Makoto},
  journal={arXiv preprint arXiv:2502.06133},
  year={2025}
}

@book{Donnelly1991,
  title     = {Quantized Vortices in Helium II},
  author    = {Donnelly, Russell J.},
  publisher = {Cambridge University Press},
  year      = {1991},
}

@book{Frisch1995,
  title     = {Turbulence: The Legacy of A.N. Kolmogorov},
  author    = {Frisch, U.},
  publisher = {Cambridge University Press},
  year      = {1995},
  doi       = {10.1017/CBO9781139170666}
}

@book{Press:CUP2007,
    author = {Press, William H. and Teukolsky, Saul A. and Vetterling, William T. and Flannery, Brian P.},
    title = {Numerical Recipes 3rd Edition: The Art of Scientific Computing},
    year = {2007},
    isbn = {0521880688},
    publisher = {Cambridge University Press}
}

@book{Pethick:CUP2008,
    title={{{{{Bose-Einstein}}}} condensation in dilute gases},
    author={Pethick, Christopher J and Smith, Henrik},
    year={2008},
    publisher={Cambridge university press}
}

@book{Pitaevskii:OUP2016,
    title={{{{{Bose-Einstein}}}} condensation and superfluidity},
    author={Pitaevskii, Lev and Stringari, Sandro},
    volume={164},
    year={2016},
    publisher={Oxford University Press}
}

@book{Barenghi:CUP2023,
    title={Quantum turbulence},
    author={Barenghi, Carlo F and Skrbek, Ladislav and Sreenivasan, Katepalli R},
    year={2023},
    publisher={Cambridge University Press},
}

@book{Tsubota:OUP2025,
    title={Quantum Hydrodynammics and Turbulence},
    author={Tsubota, Makoto and Kasamatsu, Kenichi},
    year={2025},
    publisher={Oxford University Press},
}

\end{document}